\documentclass[12pt]{article}

\usepackage[utf8]{inputenc}
\usepackage{amsmath}
\usepackage{amssymb}
\usepackage{graphicx}
\usepackage{bbold} 
\usepackage{comment}

\renewcommand{\baselinestretch}{1.2}
\newcommand{\del}{\partial}

\usepackage{braket}
\usepackage{bm}
\usepackage{bbm}
\usepackage{hyperref}
\numberwithin{equation}{section}

\newcommand{\mat}[1]{\left(\begin{matrix}#1\end{matrix}\right)}
\newcommand{\bat}[1]{\begin{matrix}#1\end{matrix}}
\newcommand{\rep}[2]{\textbf{#1}_\textbf{#2}}
\newcommand{\ihat}{\hat\imath}
\newcommand{\bs}{\boldsymbol}

\newcommand{\underlabel}[2]{\underset{#1}{\underbrace{#2}}}
\newcommand{\overlabel}[2]{\overset{#1}{\overbrace{#2}}}
\newcommand{\punc}{K}
\newcommand {\nn} {\nonumber}
\newcommand{\bbR}{{\mathbb R}}
\newcommand{\bbC}{{\mathbb C}}
\newcommand{\mycomment}[1]{}
\newcommand{\Dslash}{D\!\!\!\!/\;}
\allowdisplaybreaks

\usepackage{subfig}
\makeatother

\begin{document}
\begin{titlepage}
\renewcommand{\thefootnote}{\fnsymbol{footnote}}

\begin{flushright} 
  KEK-TH-2549
\end{flushright} 

\vspace{1.5cm}

\begin{center}
  {\bf \large A new technique to incorporate multiple fermion flavors
in tensor renormalization group method for lattice gauge theories}
\end{center}

\vspace{1cm}


\begin{center}
         Atis Y{\sc osprakob}$^{1)}$\footnote
          { E-mail address : ayosp(at)phys.sc.niigata-u.ac.jp},
         Jun N{\sc ishimura}$^{2,3)}$\footnote
          { E-mail address : jnishi(at)post.kek.jp} and
         Kouichi O{\sc kunishi}$^{1)}$\footnote
          { E-mail address : okunishi(at)phys.sc.niigata-u.ac.jp}


\vspace{1cm}

$^{1)}$\textit{Department of Physics, Niigata University,}\\
{\it Niigata 950-2181, Japan}

$^{2)}$\textit{KEK Theory Center,
Institute of Particle and Nuclear Studies,}\\
{\it High Energy Accelerator Research Organization,\\
  1-1 Oho, Tsukuba, Ibaraki 305-0801, Japan}



$^{3)}$\textit{Graduate Institute for Advanced Studies, SOKENDAI,\\
1-1 Oho, Tsukuba, Ibaraki 305-0801, Japan} 


\end{center}

\vspace{0.5cm}

\begin{abstract}
  \noindent
We propose a new technique to incorporate multiple fermion flavors in the tensor renormalization group method for lattice gauge theories, where fermions are treated by the Grassmann tensor network formalism. The basic idea is to separate the site tensor into multiple layers associated with each flavor and to introduce the gauge field in each layer as replicas, which are all identified later. This formulation, after introducing an appropriate compression scheme in the network, enables us to reduce the size of the initial tensor with high efficiency compared with a naive implementation. The usefulness of this formulation is demonstrated by investigating the chiral phase transition and the Silver Blaze phenomenon in 2D Abelian gauge theories with $N_{\rm f}$ flavors of Wilson fermions up to $N_{\rm f}=4$.
\end{abstract}
\vfill
\end{titlepage}
\vfil\eject


\renewcommand{\thefootnote}{\arabic{footnote}}
\setcounter{footnote}{0}

\section{Introduction} 
Nonperturbative computation in fermionic systems has always been challenging due to the anti-commuting nature of Grassmann variables. In Monte Carlo methods, the Grassmann variables have to be integrated out first, yielding the fermion determinant $\det M$, which makes the computation very time-consuming since the matrix $M$ has a size proportional to the system size $V$. While the computational cost can be made O($V$) by using the pseudo-fermion technique with an appropriate  Hybrid Monte Carlo algorithm, the calculation is still typically a few orders of magnitude more time-consuming than corresponding bosonic systems. Moreover, in many interesting fermionic systems such as finite density systems, strongly-correlated electron systems, and theories with chiral fermions, the fermion determinant becomes complex, which causes the notorious sign problem in conventional Monte Carlo methods. In order to overcome this problem, various methods such as the complex Langevin method \cite{Parisi:1983mgm,Nagata:2016vkn,Aarts:2009uq,Seiler:2012wz,Ito:2016efb}, the Lefschetz thimble method \cite{Picard1897,Lefschetz1924,Witten:2010cx,Witten:2010zr,Alexandru:2017oyw,Matsumoto:2021zjf,Fujisawa:2021hxh}, and the density of state method \cite{Gocksch:1987nt,Gocksch:1988iz,Langfeld:2012ah} have been developed. However, each method has its pros and cons, and many models still remain out of reach.

All these problems associated with fermionic systems can be solved beautifully in the tensor network method \cite{Okunishi:2021but,Levin:2006jai,PhysRevLett.115.180405,Hauru:2017jbf,Adachi:2020upk,PhysRevB.86.045139,Adachi:2019paf,Kadoh:2019kqk,Sakai:2017jwp,Gu:2010yh,Gu:2013gba,Shimizu:2014uva,Akiyama:2020sfo,Nakayama:2023ytr}, which is not a statistical approach based on important sampling. This method was first introduced to handle many-body systems in condensed matter physics with the main application to the calculation of the ground state based on the variational principle \cite{perezgarcia2007matrix,doi:10.1063/1.1664623,PhysRevLett.69.2863,doi:10.1143/JPSJ.65.891,doi:10.1143/JPSJ.66.3040,doi:10.1143/JPSJ.67.3066}. However, it can also be used to directly compute the partition function with some procedures based on coarse-graining, which is similar in spirits to the real-space renormalization group and hence the name, ``the tensor renormalization group (TRG) method''. Notably, it enables the computation of the partition function with a computational cost that grows only logarithmically with the system size. Although the original TRG method was proposed for a two-dimensional bosonic system \cite{Levin:2006jai}, improved versions were subsequently developed \cite{PhysRevLett.115.180405,Hauru:2017jbf,Adachi:2020upk,Nakayama:2023ytr}, and it has also been generalized to higher dimensional systems \cite{PhysRevB.86.045139,Adachi:2019paf,Kadoh:2019kqk,Sakai:2017jwp} and to fermionic systems, where Grassmann variables are treated directly \cite{Gu:2010yh,Gu:2013gba,Shimizu:2014uva,Akiyama:2020sfo,Sakai:2017jwp} unlike in Monte Carlo methods. Using this ``Grassmann tensor network'', one does not have to deal with the fermion determinant, and furthermore, the sign problem does not exist in the method from the outset because it is not a statistical approach.

A recent achievement of the TRG method is its application to gauge theory, in particular in the parameter regions that are not accessible to Monte Carlo methods due to the sign problem. Notable examples include the 2D gauge theories with a $\theta$ term \cite{Kuramashi:2019cgs,Fukuma:2021cni,Hirasawa:2021qvh}, 2D SU(2) gauge-Higgs model \cite{Bazavov:2019qih}, one-flavor Schwinger model \cite{Shimizu:2014uva,Shimizu:2014fsa,Shimizu:2017onf}, 2D QCD \cite{Bloch:2022vqz}, 3D SU(2) gauge theory \cite{Kuwahara:2022ald}, 4D $\mathbb{Z}_K$ gauge-Higgs models \cite{Akiyama:2022eip,Akiyama:2023hvt} and so on. Among these applications, gauge theories with matter fields are of particular importance since typically they are not exactly solvable. However, the TRG method has so far been applied only to the case with one fermion flavor. When there are many flavors of fermions on a single lattice site, one encounters a problem that the size of the local Hilbert space, and thus the size of the initial tensor, grows exponentially with the number of flavors. This prevents us from studying theories with multiple flavors including QCD, which is
a non-Abelian gauge theory with two (or three) flavors of light quarks.

In this paper, we propose a new technique that makes it possible to incorporate multiple flavors of fermions in gauge theory within the Grassmann tensor network formalism. The main idea is to separate the initial tensor into multiple layers associated with each flavor. Since the fermions with different flavors are interacting with the same gauge field, the interaction in the flavor direction becomes non-local after integrating out the gauge field. In order to avoid this problem, we introduce the gauge field in each layer as replicas and identify them all later. Once the system can be described by a tensor network, which is one dimension higher than the original theory due to the flavor direction, one can use the standard coarse-graining technique to compute the partition function. Our method is expected to be useful also in applying the TRG to the domain-wall formalism \cite{Kaplan:1992bt,Kaplan:1992sg} for chiral fermions regarding the flavor direction in our method as the extra space-time dimension.
We also introduce an efficient compression scheme that further reduces the size of the initial tensor drastically, especially at large $K$. For $N_{\rm f}=1$, the performance of our method is found to be as good as in the previous calculations for the Schwinger model \cite{Shimizu:2014uva,Shimizu:2014fsa,Shimizu:2017onf}.

The usefulness of this formulation is demonstrated by applying it to Abelian gauge theories in two dimensions with $N_{\rm f}$ flavors of Wilson fermions.
First, we investigate the chiral phase transition in the $N_{\rm f}=2$ case, which shows that the result obtained in the $\mathbb{Z}_K$ gauge theory converges to the $\text{U}(1)$ result obtained by using the Monte Carlo method \cite{Hip:1997em} as we increase $K$.
Next, we investigate the Silver Blaze phenomenon in the case of finite fermion density up to $N_{\rm f}=4$.

While this paper was being completed, we encountered a paper \cite{Akiyama:2023lvr} which addresses the same issue of treating multiple fermion flavors in the TRG method. There the initial tensor for all the flavors is separated into multiple layers by using the matrix product decomposition, which requires the memory of order O($e^{c N_{\rm f}}$). In contrast, the memory cost of our method is of order O(1) since the layers are separated analytically and all the layers are identical. This memory cost reduction enabled the investigation of gauge theories, which was not possible in Ref.~\cite{Akiyama:2023lvr}.

The rest of this paper is organized as follows. In section \ref{section:formulation}, we explain our basic idea to implement multiple flavors in the TRG method. In particular, we derive the initial tensor for 2D Abelian gauge theories with Wilson fermions as an example. In section \ref{section:procedures}, we describe how we perform the procedures of the TRG method using the initial tensor. In particular, we discuss how we compress the initial tensor efficiently by inserting isometries and explain how we perform coarse-graining in the flavor space. In section \ref{section:results}, we present the numerical results obtained by our method. After showing some results of the performance tests concerning the initial tensor compression and the coarse-graining procedure in the flavor direction, we demonstrate the usefulness of our method by investigating the chiral phase transition and the Silver Blaze phenomenon in 2D Abelian gauge theories. Section \ref{section:summary} is devoted to a summary and discussions. In Appendix \ref{section:grassmann_review}, we give a brief review of the Grassmann tensor network. In Appendix \ref{section:algorithm}, we describe the coarse-graining algorithm in detail. In Appendix \ref{section:interflavor_interaction}, we discuss the generalization of our technique to a model with local multi-flavor interactions, which is important, in particular, in applying our method to the domain-wall formalism.

\section{The basic idea to implement multiple flavors}
\label{section:formulation}

In this section, we explain our basic idea
to implement multiple flavors in the TRG method.
First, we split the system into multiple layers
by introducing replicas of gauge fields for each flavor,
and then we construct the initial tensor
based on the Grassmann tensor network formalism
including the flavor direction.

\subsection{Splitting the system into multiple layers}

For simplicity, we will describe our idea in the case of 
Abelian gauge theory $G\subseteq\text{U}(1)$
on a two-dimensional square lattice $\Lambda_2$ with the lattice spacing $a$
although it can be readily applied to higher dimensions and non-Abelian cases.
The gauge field $A_{x,\mu}$ is represented on the lattice by the
link variable $U_{x,\mu}=\exp(iaA_{x,\mu})\equiv\exp(i\varphi_{x,\mu})\in G$.
Let us then consider the lattice action with $N_{\rm f}$ flavors of
Wilson fermions
given by
\begin{align}
    S&=S_\text{gauge}[\varphi]+\sum_{x\in\Lambda_2}\sum_{\alpha=1}^{N_{\rm f}}\bar{\psi}_{x}^{(\alpha)}\Dslash^{(\alpha)}\psi_{x}^{(\alpha)} \ , \\
    S_\text{gauge}[\varphi]&=\beta\sum_{x\in\Lambda_2}\left\{1-\cos(\varphi_{x,1}+\varphi_{x+\hat 1,2}-\varphi_{x+\hat 2,1}-\varphi_{x,2})\right\} \  ,\\
    \Dslash^{(\alpha)}\psi_{x}^{(\alpha)} 
    &=-\frac{1}{2}\sum_{\nu=1,2}((\mathbb{1}-\gamma_\nu )e^{+(\tilde\mu_\alpha\delta_{\nu,2}+iq_\alpha\varphi_{x,\nu})}\psi_{x+\hat\nu}^{(\alpha)}
    +(\mathbb{1}+\gamma_\nu )e^{-(\tilde\mu_\alpha\delta_{\nu,2}+iq_\alpha\varphi_{x-\hat\nu,\nu})}\psi_{x-\hat\nu}^{(\alpha)})\nonumber\\
    &\qquad\qquad+ \left(\tilde m_\alpha+2\right)\psi_{x}^{(\alpha)} \ ,
    \label{fermion-action}
\end{align}
where
$\beta=1/(ga)^2$ is the inverse gauge coupling and
and $\gamma_\nu$ are the 2D gamma matrices given, for instance,
by the Pauli matrices $\gamma_1=\sigma_1$ and $\gamma_2=\sigma_2$.
The fermion fields are represented 
by
the 2-component Dirac spinors
$\psi_x^{(\alpha)}$ for each flavor $\alpha$ with charge $q_\alpha$.
We also introduce
dimensionless chemical potential $\tilde\mu_\alpha$
and mass $\tilde m_\alpha$ for each flavor $\alpha$.
%

Since fermions have $4N_{\rm f}$ internal degrees of freedom,
the local Hilbert space at a given lattice site
has dimension $D=2^{4N_{\rm f}}$. Consequently, the size of the initial tensor grows as $D^4=2^{16N_{\rm f}}$. It is thus beneficial to separate different flavors
from each other to avoid the exponential growth of the tensor size with $N_{\rm f}$.
To that end,
we split the link variables into $N_{\rm f}$ replicas
and define the partition function as
\begin{equation}
    Z=\int D\varphi \prod_{\alpha=1}^{N_{\rm f}}\left(D\varphi^{(\alpha)}D\psi^{(\alpha)} D\bar\psi^{(\alpha)}\right) \delta(\varphi^{(\alpha)}-\varphi)e^{-\sum_\alpha S^{(\alpha)}} \ ,
\end{equation}
where the new action $S^{(\alpha)}$ is given as
\begin{equation}
    S^{(\alpha)}=\frac{1}{N_{\rm f}}S_\text{gauge}[\varphi^{(\alpha)}]+\sum_{x\in\Lambda_2}\bar{\psi}_{x}^{(\alpha)}\Dslash^{(\alpha)}\psi_{x}^{(\alpha)} \ .
\end{equation}
This enables us to
split the Boltzmann weight into several layers corresponding to each replica,
which are linked by the delta function.
This decomposition into multiple layers has recently been considered
in Ref.~\cite{Akiyama:2023lvr} but with the matrix product decomposition instead
of using the delta function.
Let us emphasize, however, that separating the layers by hand from the beginning
as we do here is crucial in avoiding completely
the singular value decomposition, which can be both
memory-consuming and computationally expensive in lattice gauge theories.

While we have not introduced local interactions among different flavors,
our tensor construction can be generalized to such cases
as described in Appendix \ref{section:interflavor_interaction}.

\begin{figure}
    \centering
    \includegraphics[scale=0.8]{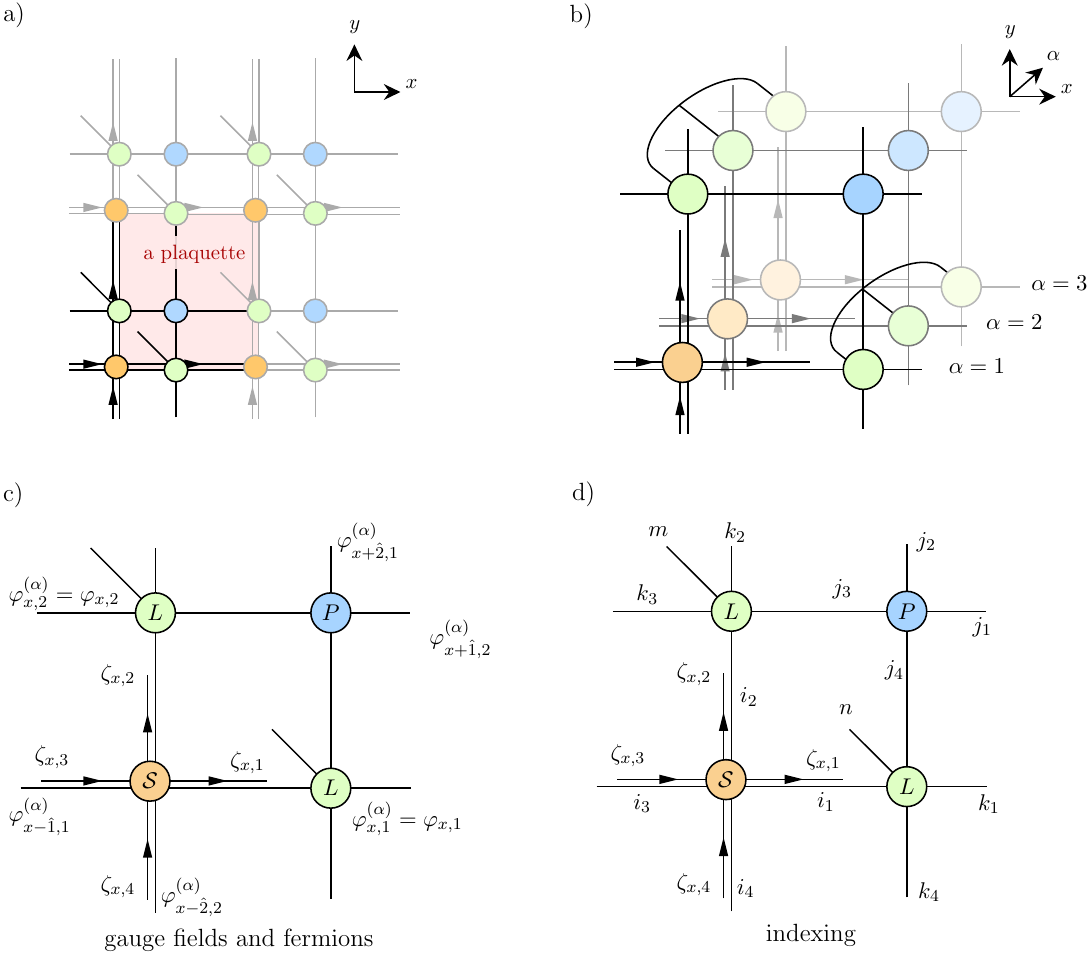}
    \caption{a) The two-dimensional tiling of the site tensor, which
      is composed of four subtensors. The shaded region represents a plaquette.
      b) The three-dimensional tiling of the site tensor for $N_{\rm f}=3$. Each layer corresponds to some flavor $\alpha$. c) The site tensor \eqref{eq:site_tensor_original} with the gauge fields and fermionic variables. d) The site tensor with the index for each bond. In the diagrams, fermionic legs with an arrow pointing away from the tensor are non-conjugated fermions,
      while those with an arrow pointing into the tensor are conjugated fermions.}
    \label{fig:site_tensor}
\end{figure}

\subsection{Constructing the tensor network}

We treat the fermion fields by the Grassmann tensor
network \cite{Akiyama:2020sfo} (See Appendix \ref{section:grassmann_review}
for the details.).
%
Let us rewrite the fermion action in the following form
\begin{align}
\bar{\psi}_{x}^{(\alpha)}\Dslash^{(\alpha)}\psi_{x}^{(\alpha)} 
  &=\bar{\psi}_{x}^{(\alpha)} W_{x}^{(\alpha)}\psi_{x}^{(\alpha)}+\sum_\nu\left(\bar{\psi}_{x}^{(\alpha)} H_{x,+\nu}^{(\alpha)}\psi_{x+\hat\nu}^{(\alpha)}+\bar{\psi}_{x}^{(\alpha)} H_{x,-\nu}^{(\alpha)}\psi_{x-\hat\nu}^{(\alpha)}\right) \ ,\\
    W_{x}^{(\alpha)}&=\tilde m_\alpha+2 \ ,\\
    H_{x,+\nu}^{(\alpha)}&=-\frac{1}{2}(\mathbb{1}-\gamma_\nu)\, e^{+(\tilde\mu_\alpha\delta_{\nu,2}+iq_\alpha\varphi_{x,\nu}^{(\alpha)})} \ ,\\
    H_{x,-\nu}^{(\alpha)}&=-\frac{1}{2}(\mathbb{1}+\gamma_\nu)\, e^{-(\tilde\mu_\alpha\delta_{\nu,2}+iq_\alpha\varphi_{x-\hat\nu,\nu}^{(\alpha)})} \ .
\end{align}
Here we transform the site fermions $(\psi,\bar\psi)$ into auxiliary link fermions $(\eta,\bar\eta)$
by using the relation\footnote{This follows from the identity
$$
  e^{-h\bar\theta\theta}=\int d\bar\eta \, d\eta \,
  e^{-\bar\eta\eta-\bar\theta\eta+h\bar\eta\theta}
$$
  for a Grassmann-even constant $h$ and one-component Grassmann-odd numbers $\theta,\bar\theta,\eta,\bar\eta$, which can be generalized to
  the multi-component case in a straightforward manner.}

\begin{align}
    e^{-\bar{\psi}_{x}^{(\alpha)}H_{x,\pm\nu}^{(\alpha)}\psi_{x\pm\hat\nu}^{(\alpha)}}&=\int d\bar{\eta}_{x,\pm\nu}^{(\alpha)}d\eta_{x,\pm\nu}^{(\alpha)}
    e^{-\bar{\eta}_{x,\pm\nu}^{(\alpha)}\eta_{x,\pm\nu}^{(\alpha)}-\bar{\psi}_{x}^{(\alpha)}\eta_{x,\pm\nu}^{(\alpha)}+\bar{\eta}_{x,\pm\nu}^{(\alpha)}H_{x,\pm\nu}^{(\alpha)}\psi_{x\pm\hat\nu}^{(\alpha)}} \ .
\end{align}

The integration of the link variables
is performed by the summation
\begin{equation}
  \int_{\mathbb{Z}_K}d\varphi f(\varphi)\equiv\sum_{k=1}^{K} wf(\varphi_k)
    \label{eq:ZKmeasure}
\end{equation}
in the case of $\mathbb{Z}_K$ gauge theory,
where $w=1/2\pi K$ and $\varphi_k=2(k-1)\pi/K$.
In the case of U(1) gauge theory, we approximate
the group integral by the Gaussian quadrature \cite{Kuramashi:2019cgs} as
\begin{equation}
    \int_\text{U(1)}d\varphi f(\varphi)\equiv\int_{-\pi}^{+\pi}\frac{d\varphi}{2\pi}f(\varphi)\approx\sum_{k=1}^Kw(\varphi_k)f(\varphi_k) \ ,
\end{equation}
where the weight function $w(\varphi)$ and the nodes $\varphi_k$
depend on the quadrature.

Thus we arrive at
\begin{equation}
    Z=\int_{\bar{\eta}\eta} \sum_{\{\varphi\}} \prod_{x,\alpha}\mathcal{T}_{x}^{(\alpha)} \ ,
\end{equation}
where we have defined the tensor
\begin{align}
    \mathcal{T}_{x}^{(\alpha)}&=P^{(\alpha)}_{x} \mathcal{S}^{(\alpha)}_{x}L^{(\alpha)}_{x,1}L^{(\alpha)}_{x,2} \ ,
    \label{eq:site_tensor_original}\\
    P^{(\alpha)}_{x}&=\left\{w(\varphi^{(\alpha)}_{x,1})w(\varphi^{(\alpha)}_{x,2})\right\}^{1/N_{\rm f}}e^{\frac{\beta}{N_{\rm f}}\cos(\varphi^{(\alpha)}_{x,1}+\varphi^{(\alpha)}_{x+\hat 1,2}-\varphi^{(\alpha)}_{x+\hat 2,1}-\varphi^{(\alpha)}_{x,2})}
    \label{eq:site_tensor_P} \ ,\\
    L^{(\alpha)}_{x,\mu}&=\delta(\varphi^{(\alpha)}_{x,\mu}-\varphi_{x,\mu}).\label{eq:site_tensor_L}\\
    \mathcal{S}^{(\alpha)}_{x}&=
    \int d\psi_{x}^{(\alpha)}d\bar{\psi}_{x}^{(\alpha)}\exp\left[-\bar{\psi}_{x}^{(\alpha)} W_{x}^{(\alpha)}\psi_{x}^{(\alpha)}
    -\sum_{\pm,\nu}
    \left\{\bar{\psi}_{x}^{(\alpha)}\eta_{x,\pm\nu}^{(\alpha)}-\bar{\eta}_{x\mp\hat\nu,\pm\nu}^{(\alpha)}H_{x\mp\hat\nu,\pm\nu}^{(\alpha)}\psi_{x}^{(\alpha)}\right\}\right] \ ,
    \label{eq:site_tensor_S}
\end{align}
and introduced a short-hand notation for the Grassmann integral
\begin{equation}
    \int_{\bar{\eta}\eta}\equiv\int\prod_{x,\nu,\alpha}\left(d\bar{\eta}_{x,\nu}^{(\alpha)} d\eta_{x,\nu}^{(\alpha)}e^{-\bar{\eta}_{x,\nu}^{(\alpha)}\eta_{x,\nu}^{(\alpha)}}\right) \ .
\end{equation}
Note that the tensors $P$, $L$ and $\mathcal{S}$
are associated with the plaquettes, links, and sites, respectively.
The connection of these tensors is shown
in Fig.~\ref{fig:site_tensor}-c).

Performing the integral \eqref{eq:site_tensor_S} symbolically\footnote{We use Mathematica v13.1.0.0 with a package for non-commutative algebra, NCAlgebra v5.0.6.},
we obtain the tensor $\mathcal{S}$
in the form of a polynomial of link fermions given as
\begin{align}
    \mathcal{S}^{(\alpha)}_x=\sum_{\{I,J\}}
    &(C^{(\alpha)}_x)_{I_1J_1I_2J_2I_3J_3I_4J_4}(\varphi^{(\alpha)}_{x,1},\varphi^{(\alpha)}_{x,2},\varphi^{(\alpha)}_{x-\hat 1,1},\varphi^{(\alpha)}_{x-\hat 2,2})
    \\
    &\qquad\times\eta_{x,+1}^{I_1}\bar\eta_{x+\hat 1,-\hat 1}^{J_1}
    \eta_{x,+2}^{I_2}\bar\eta_{x+\hat 2,-\hat 2}^{J_2}
    \bar\eta_{x-\hat 1,+\hat 1}^{I_3}\eta_{x,-1}^{J_3}
    \bar\eta_{x-\hat 2,+\hat 2}^{I_4}\eta_{x,-2}^{J_4} \ ,
    \nonumber
\end{align}
where the coefficient $C_x^{(\alpha)}$
depends on
the gauge link variables.
Here we have introduced 
\begin{align}
    \eta^I&\equiv\theta_1^{k_1}\theta_2^{k_2},\\
    I&\equiv(k_1,k_2) \ ,
    \label{eq:multi_index2}
\end{align}
where $\theta_1$ and $\theta_2$ are the two components of $\eta$
and $k_1,k_2\in\{0,1\}$ represent the occupation number of the components.
Since some of these link fermions connect the same pair of sites,
it is convenient to combine the fermion indices
as $(I_a,J_a)\mapsto K_a$ with the prescription
\begin{align}
  \zeta_{x,1}^{K_1}&=\eta_{x,+1}^{I_1}\bar\eta_{x+\hat 1,-\hat 1}^{J_1} \ ,
  \label{eq:Psix1}\\
  \zeta_{x,2}^{K_2}&=\eta_{x,+2}^{I_2}\bar\eta_{x+\hat 2,-\hat 2}^{J_2} \ ,
  \label{eq:Psix2}\\
  \bar\zeta_{x,3}^{K_3}&=(-)^{p(J_3)}\bar\eta_{x-\hat 1,+\hat 1}^{I_3}\eta_{x,-1}^{J_3} \ ,
  \label{eq:Psix3}\\
  \bar\zeta_{x,4}^{K_4}&=(-)^{p(J_4)}\bar\eta_{x-\hat 2,+\hat 2}^{I_4}\eta_{x,-2}^{J_4} \ ,
  \label{eq:Psix4}
\end{align}
where $p(J)$ is the Grassmann parity of the Grassmannn number $\eta^J$ defined by
\begin{equation}
  p(J)=\sum_aj_a
  \label{eq:parity_defninition}
\end{equation}
with $j_a$ being the fermion occupation number of the $a$-th component.
The sign factor $(-)^{p(J)}$ is introduced for the consistency
of Grassmann tensor contraction (See \eqref{eq:join_leg1}-\eqref{eq:join_leg2}.).

Using the Grassmann index notation \eqref{eq:Grassmann_tensor_expansion_original},
the
tensor $\mathcal{S}$ can be expanded as
(omitting the site index $x$ to avoid redundancy)
\begin{align}
  \mathcal{S}^{(\alpha)}_{\zeta_1\zeta_2\bar\zeta_3\bar\zeta_4;i_1i_2i_3i_4}
  &=\sum_{\{K\}}S^{(\alpha)}_{K_1K_2K_3K_4;i_1i_2i_3i_4}\zeta_1^{K_1}\zeta_2^{K_2}\bar\zeta_3^{K_3}\bar\zeta_4^{K_4} \ ,
  \label{eq:Btensor}\\
  S^{(\alpha)}_{K_1K_2K_3K_4;i_1i_2i_3i_4}
  &=C^{(\alpha)}_{I_1J_1I_2J_2I_3J_3I_4J_4}(\varphi^{(\alpha)}_{i_1},\varphi^{(\alpha)}_{i_2},\varphi^{(\alpha)}_{i_3},\varphi^{(\alpha)}_{i_4})(-)^{p(J_1)+p(J_2)}
\end{align}
with the sign factors given in \eqref{eq:Psix3} and \eqref{eq:Psix4}.
Here the index $i$ in $\varphi^{(\alpha)}_i$ refers to the index
of the quadrature node in \eqref{eq:ZKmeasure}.
The site and orientation indices of the link variables
$\varphi_{x,\mu}$
are omitted.

The tensor $P$ can be rewritten in terms of the quadrature indices as
\begin{equation}
    P^{(\alpha)}_{j_1j_2j_3j_4}=\left\{w(\varphi^{(\alpha)}_{j_3})w(\varphi^{(\alpha)}_{j_4})\right\}^{1/N_{\rm f}}e^{\frac{\beta}{N_{\rm f}}\cos(\varphi^{(\alpha)}_{j_4}+\varphi^{(\alpha)}_{j_1}-\varphi^{(\alpha)}_{j_2}-\varphi^{(\alpha)}_{j_3})} \ .
\end{equation}
The tensor $L$ \eqref{eq:site_tensor_L},
which depends on two fields, 
has actually five legs because it is connected to
two plaquettes, two sites,
and to the global gauge field $\varphi_{x,\mu}$.
Therefore, it can be written in terms of the quadrature indices as
\begin{equation}
    L_{i_\mu i_\nu k_\mu k_\nu m}=\delta_{mi_\mu}\delta_{mi_\nu}\delta_{mk_\mu}\delta_{mk_\nu} \ .
\end{equation}

To summarize,
the coefficient of the site tensor \eqref{eq:site_tensor_original}
is given by
\begin{equation}
    T^{(\alpha)}_{I_1I_2I_3I_4;j_1k_1j_2k_2i_3k_3i_4k_4;mn}=\sum_{i_1,i_2,j_3,j_4}S^{(\alpha)}_{I_1I_2I_3I_4;i_1i_2i_3i_4}P^{(\alpha)}_{j_1j_2j_3j_4} L_{i_2j_3k_2k_3m} L_{j_4i_1k_4k_1n} \ .
    \label{eq:site_tensor}
\end{equation}
In the above expression, the indices with subscripts 1, 2, 3, and 4
are associated with the legs pointing
in the direction $+\hat1$, $+\hat2$, $-\hat1$ and $-\hat2$, respectively,
whereas $m$ and $n$ are associated with $\varphi_{x,2}$ and $\varphi_{x,1}$,
respectively.
The schematic representation of the site tensor is given in
Fig.~\ref{fig:site_tensor}-d).

\begin{figure}
    \centering
    \includegraphics[scale=0.8]{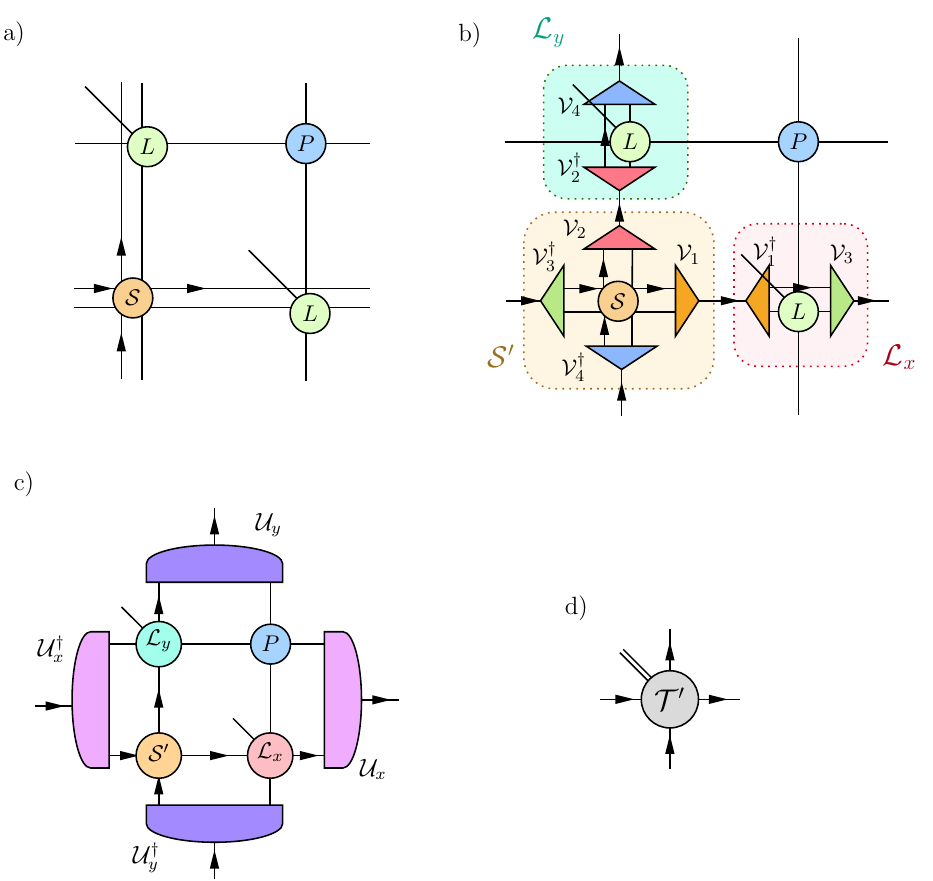}
    \caption{The schematic representation of the initial tensor compression.
      a) The original tensor.
      b) Four sets of isometries are inserted between $L$'s and $\mathcal{S}$.
      c) Another set of isometries is applied to compress the whole tensor.
      d) The compressed initial tensor.
      The tensor $\mathcal{V}_a$ in the diagram above is the isometry with the fastest-falling singular value between $\mathcal{V}_{+a}$ and $\mathcal{V}_{-a}$. Similarly, $\mathcal{U}_\mu$ is the isometry with the fastest-falling singular value between $\mathcal{U}_{+\mu}$ and $\mathcal{U}_{-\mu}$.
      }
    \label{fig:compactification}
\end{figure}

\section{The procedures of the TRG}
\label{section:procedures}

In this section, we describe how we perform the
procedures of the TRG method using the initial tensor derived in
the previous section.
In particular, we discuss how we compress the initial tensor efficiently by inserting isometries and explain how we perform coarse-graining in the flavor space.

\subsection{Compressing the initial tensor}
\label{section:compression}

As one can see from the expression \eqref{eq:site_tensor},
the initial tensor for lattice gauge theories has typically
a large dimension due to the existence of many legs.
%
It is therefore
important to compress its size first before we perform
the coarse-graining procedure.
Here we use the compressing procedure
based on the higher-order SVD,
which is frequently used in
HOTRG-type algorithms \cite{doi:10.1137/S0895479896305696, PhysRevB.86.045139} (See also Appendix \ref{section:HOTRG}.).

\begin{figure}
    \centering
    \includegraphics[scale=0.9]{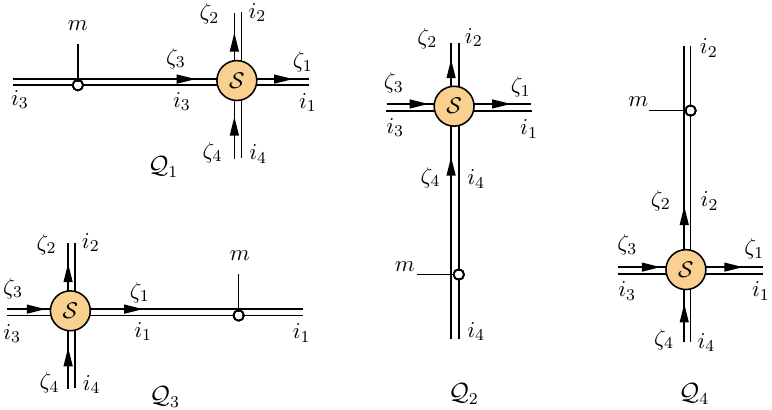}
    \caption{The $\mathcal{Q}$ tensors used in the computation of the four isometries. $\mathcal{Q}_1$ and $\mathcal{Q}_2$ are used to compute the inner isometries,
      while $\mathcal{Q}_3$ and $\mathcal{Q}_4$ are used to compute the outer isometries in Fig.~\ref{fig:compactification}-b).
      The five-legged $L$-tensor in
      Fig.~\ref{fig:compactification}-a)
      are replaced by the three-legged nodes since both of these diagrams result in the same $M$ matrices \eqref{eq:Mfirst}-\eqref{eq:Mlast}.
      Note that the position of $\mathcal{S}$ and the Kronecker delta nodes
      are swapped in the inner and outer cases for a given axis.}
    \label{fig:QB}
\end{figure}

The first step of the compressing procedure
is to ``squeeze'' the legs of the $\mathcal{S}$ tensor $(\zeta_a,i_a)$
into a smaller leg $\xi_a$ using the hybrid isometries
that merge a fermionic leg and a bosonic leg into one fermionic leg as
\begin{align}
    \mathcal{S}^{(\alpha)'}_{\xi_1\xi_2\bar\xi_3\bar\xi_4}
    &
    =
    \int_{\{\bar\zeta_\nu\zeta_\nu\}}
    \sum_{\{i_\nu\}}
    (\mathcal{V}_4)^\dagger_{(\bar\xi_4)(\zeta_4i_4)}
    (\mathcal{V}_3)^\dagger_{(\bar\xi_3)(\zeta_3i_3)}
    \mathcal{S}^{(\alpha)}_{\zeta_1\zeta_2\bar\zeta_3\bar\zeta_4;i_1i_2i_3i_4}
    (\mathcal{V}_1)_{(\bar\zeta_1i_1)(\xi_1)}
    (\mathcal{V}_2)_{(\bar\zeta_2i_2)(\xi_2)} \ ,
    \label{eq:Bprime}
\end{align}
where the Hermitian conjugate is defined in \eqref{eq:conjugate}
and 
the contraction $\int_{\bar\zeta_\nu\zeta_\nu}$ of the bond $(\zeta_\nu,\bar\zeta_\nu)$
is defined in \eqref{eq:contraction_definition}.
Here we have inserted four isometries; namely
$\mathcal{V}_1$ and $\mathcal{V}_2$ are inserted in
the inner links between $\mathcal{S}$ and the Kronecker delta nodes $L$,
whereas $\mathcal{V}_3$ and $\mathcal{V}_4$ are inserted in
the outer links (See Fig.~\ref{fig:compactification}-b).
In order to obtain these isometries,
we first define the tensors (repeating indices are not summed; See Fig.~\ref{fig:QB}.)
\begin{align}
    (\mathcal{Q}_1)_{\zeta_1\zeta_2\bar\zeta_3\bar\zeta_4;i_1i_2i_3i_4;m}&=\mathcal{S}^{(\alpha)}_{\zeta_1\zeta_2\bar\zeta_3\bar\zeta_4;i_1i_2i_3i_4}\delta_{i_3m} \ ,\\
    (\mathcal{Q}_2)_{\zeta_1\zeta_2\bar\zeta_3\bar\zeta_4;i_1i_2i_3i_4;m}&=\mathcal{S}^{(\alpha)}_{\zeta_1\zeta_2\bar\zeta_3\bar\zeta_4;i_1i_2i_3i_4}\delta_{i_4m}\ ,\\
    (\mathcal{Q}_3)_{\zeta_1\zeta_2\bar\zeta_3\bar\zeta_4;i_1i_2i_3i_4;m}&=\mathcal{S}^{(\alpha)}_{\zeta_1\zeta_2\bar\zeta_3\bar\zeta_4;i_1i_2i_3i_4}\delta_{i_1m}\ ,\\
    (\mathcal{Q}_4)_{\zeta_1\zeta_2\bar\zeta_3\bar\zeta_4;i_1i_2i_3i_4;m}&=\mathcal{S}^{(\alpha)}_{\zeta_1\zeta_2\bar\zeta_3\bar\zeta_4;i_1i_2i_3i_4}\delta_{i_2m} 
\end{align}
and construct the Grassmann Hermitian matrices
\begin{align}
    (\mathcal{M}_{+1})_{(\bar\zeta_1 i_1)(\zeta_1' i_1')}&=\!\!\!\!
    \sum_{i_2,i_3,i_4,m}\int_{\substack{\bar\zeta_2\zeta_2,\\\bar\zeta_3\zeta_3,\\\bar\zeta_4\zeta_4}}
    (\mathcal{Q}_1)^\dagger_{(\bar\zeta_1 i_1)(\bar\zeta_2\zeta_3\zeta_4i_2i_3i_4m)}(\mathcal{Q}_1)_{(\zeta_2\bar\zeta_3\bar\zeta_4i_2i_3i_4m)(\zeta_1' i_1')} \ ,\label{eq:Mfirst}\\
    (\mathcal{M}_{-1})_{(\bar\zeta_3 i_3)(\zeta_3' i_3')}&=\!\!\!\!
    \sum_{i_1,i_2,i_4,m}\int_{\substack{\bar\zeta_1\zeta_1,\\\bar\zeta_2\zeta_2,\\\bar\zeta_4\zeta_4}}
    (\mathcal{Q}_1)_{(\bar\zeta_3 i_3)(\zeta_1\zeta_2\bar\zeta_4i_1i_2i_4m)}(\mathcal{Q}_1)^\dagger_{(\bar\zeta_1\bar\zeta_2\zeta_4i_1i_2i_4m)(\zeta_3' i_3')} \ ,\\
    (\mathcal{M}_{+2})_{(\bar\zeta_2 i_2)(\zeta_2' i_2')}&=\!\!\!\!
    \sum_{i_1,i_3,i_4,m}\int_{\substack{\bar\zeta_1\zeta_1,\\\bar\zeta_3\zeta_3,\\\bar\zeta_4\zeta_4}}
    (\mathcal{Q}_2)^\dagger_{(\bar\zeta_2 i_2)(\bar\zeta_1\zeta_3\zeta_4i_1i_3i_4m)}(\mathcal{Q}_2)_{(\zeta_1\bar\zeta_3\bar\zeta_4i_1i_3i_4m)(\zeta_2' i_2')} \ ,\\
    (\mathcal{M}_{-2})_{(\bar\zeta_4 i_4)(\zeta_4' i_4')}&=\!\!\!\!
    \sum_{i_1,i_2,i_3,m}\int_{\substack{\bar\zeta_1\zeta_1,\\\bar\zeta_2\zeta_2,\\\bar\zeta_3\zeta_3}}
    (\mathcal{Q}_2)_{(\bar\zeta_4 i_4)(\zeta_1\zeta_2\bar\zeta_3i_1i_2i_3m)}(\mathcal{Q}_2)^\dagger_{(\bar\zeta_1\bar\zeta_2\zeta_3i_1i_2i_3m)(\zeta_4' i_4')} \ ,\\
    (\mathcal{M}_{+3})_{(\bar\zeta_1 i_1)(\zeta_1' i_1')}&=\!\!\!\!
    \sum_{i_2,i_3,i_4,m}\int_{\substack{\bar\zeta_2\zeta_2,\\\bar\zeta_3\zeta_3,\\\bar\zeta_4\zeta_4}}
    (\mathcal{Q}_3)^\dagger_{(\bar\zeta_1 i_1)(\bar\zeta_2\zeta_3\zeta_4i_2i_3i_4m)}(\mathcal{Q}_3)_{(\zeta_2\bar\zeta_3\bar\zeta_4i_2i_3i_4m)(\zeta_1' i_1')} \ ,\\
    (\mathcal{M}_{-3})_{(\bar\zeta_3 i_3)(\zeta_3' i_3')}&=\!\!\!\!
    \sum_{i_1,i_2,i_4,m}\int_{\substack{\bar\zeta_1\zeta_1,\\\bar\zeta_2\zeta_2,\\\bar\zeta_4\zeta_4}}
    (\mathcal{Q}_3)_{(\bar\zeta_3 i_3)(\zeta_1\zeta_2\bar\zeta_4i_1i_2i_4m)}(\mathcal{Q}_3)^\dagger_{(\bar\zeta_1\bar\zeta_2\zeta_4i_1i_2i_4m)(\zeta_3' i_3')}\ ,\\
    (\mathcal{M}_{+4})_{(\bar\zeta_2 i_2)(\zeta_2' i_2')}&=\!\!\!\!
    \sum_{i_1,i_3,i_4,m}\int_{\substack{\bar\zeta_1\zeta_1,\\\bar\zeta_3\zeta_3,\\\bar\zeta_4\zeta_4}}
    (\mathcal{Q}_4)^\dagger_{(\bar\zeta_2 i_2)(\bar\zeta_1\zeta_3\zeta_4i_1i_3i_4m)}(\mathcal{Q}_4)_{(\zeta_1\bar\zeta_3\bar\zeta_4i_1i_3i_4m)(\zeta_2' i_2')} \ ,\\
    (\mathcal{M}_{-4})_{(\bar\zeta_4 i_4)(\zeta_4' i_4')}&=\!\!\!\!
    \sum_{i_1,i_2,i_3,m}\int_{\substack{\bar\zeta_1\zeta_1,\\\bar\zeta_2\zeta_2,\\\bar\zeta_3\zeta_3}}
    (\mathcal{Q}_4)_{(\bar\zeta_4 i_4)(\zeta_1\zeta_2\bar\zeta_3i_1i_2i_3m)}(\mathcal{Q}_4)^\dagger_{(\bar\zeta_1\bar\zeta_2\zeta_3i_1i_2i_3m)(\zeta_4' i_4')} \ ,
    \label{eq:Mlast}
\end{align}
where the indices in the parenthesis are combined into a single index
with the prescription \eqref{eq:join_leg1}-\eqref{eq:join_leg2},
and the Hermitian conjugate of a two-legged Grassmann tensor is defined
in \eqref{eq:conjugate}.
Note also
that the coefficient tensor with
reordered indices should have
appropriate sign factors due to the permutation of Grassmann-odd variables.

Then we diagonalize the Hermitian matrices $\mathcal{M}_{\pm a}$,
which gives the unitary matrices $\mathcal{V}_{\pm a}$.
By comparing the singular value spectra
of $\mathcal{M}_{+a}$ and $\mathcal{M}_{-a}$,
we define the isometry in \eqref{eq:Bprime}
by the unitary matrix $\mathcal{V}_{\pm a}$
that corresponds to the one with the fastest falling spectrum.


Corresponding to \eqref{eq:Bprime},
we have to attach the same isometries on $L$ as
\begin{align}
    (\mathcal{L}^{(\alpha)}_x)_{\bar\xi_1\xi_3j_4k_4n}
    &=\sum_{i_1,k_1}\int_{\bar\zeta_x\zeta_x}(\mathcal{V}_1)^\dagger_{(\bar \xi_1)(\zeta_xi_1)}(\mathcal{V}_3)_{(\bar\zeta_xk_1)(\xi_3)}L_{j_4i_1k_4k_1n}\\
    &=\int_{\bar\zeta_x\zeta_x}(\mathcal{V}_1)^\dagger_{(\bar \xi_1)(\zeta_xn)}(\mathcal{V}_3)_{(\bar\zeta_xn)(\xi_3)}\delta_{j_4n}\delta_{k_4n} \ ,\\
    (\mathcal{L}^{(\alpha)}_y)_{\bar\xi_2\xi_4j_3k_3m}
    &=\sum_{i_2,k_2}\int_{\bar\zeta_y\zeta_y}(\mathcal{V}_2)^\dagger_{(\bar \xi_2)(\zeta_yi_2)}(\mathcal{V}_4)_{(\bar\zeta_yk_2)(\xi_4)}L_{i_2j_3k_2k_3m}\\
    &=\int_{\bar\zeta_y\zeta_y}(\mathcal{V}_2)^\dagger_{(\bar \xi_2)(\zeta_ym)}(\mathcal{V}_4)_{(\bar\zeta_ym)(\xi_4)}\delta_{j_3m}\delta_{k_3m} \ .
\end{align}
Despite having many indices,
these tensors are actually sparse due to the Kronecker deltas,
which
makes it more efficient to use sparse array algorithms to perform the calculation.
The schematic representation
of the construction of $\mathcal{S}'$, $\mathcal{L}_x$ and $\mathcal{L}_y$
is given in Fig.~\ref{fig:compactification}-b.

Now that we have rewritten the site tensors in terms of four sparse subtensors,
we can proceed to perform the final compression,
which further reduces the size of the site tensor.
This can be done in exactly the same way as we have done
in compressing $\mathcal{S}$. We first contract all the tensors together as
\begin{align}
    \tilde{\mathcal{T}}^{(\alpha)}_{\xi_1\xi_2\bar\xi_3\bar\xi_4;i_1i_2i_3i_4;mn}
    &=
    \int_{\bar\xi_x\xi_x,\bar\xi_y\xi_y}
    P^{(\alpha)}_{i_1i_2i_3i_4}
    \mathcal{S}^{(\alpha)'}_{\xi_x\xi_y\bar\xi_3\bar\xi_4}
    (\mathcal{L}^{(\alpha)}_x)_{\bar\xi_x\xi_1i_4i_4n}
    (\mathcal{L}^{(\alpha)}_y)_{\bar\xi_y\xi_2i_3i_3m}
     \ ,
    \label{eq:ABcontraction}
\end{align}
where the repeated indices $i_3$ and $i_4$ are not summed over.
Then we construct the Hermitian matrices
\begin{align}
    (\tilde{\mathcal{M}}_{+x})_{(\bar\xi_1 i_1)(\xi_1' i_1')}&=\!\!\!\!
    \sum_{i_2,i_3,i_4,m,n}\int_{\bar\xi_2\xi_2,\bar\xi_3\xi_3,\bar\xi_4\xi_4}
    \tilde{\mathcal{T}}^{(\alpha)\dagger}_{(\bar\xi_1 i_1)(\bar\xi_2\xi_3\xi_4i_2i_3i_4mn)}\tilde{\mathcal{T}}^{(\alpha)}_{(\xi_2\bar\xi_3\bar\xi_4i_2i_3i_4mn)(\xi_1' i_1')} \ ,
    \label{eq:Mtildefirst}\\
    (\tilde{\mathcal{M}}_{-x})_{(\bar\xi_3 i_3)(\xi_3' i_3')}&=\!\!\!\!
    \sum_{i_1,i_2,i_4,m,n}\int_{\bar\xi_1\xi_1,\bar\xi_2\xi_2,\bar\xi_4\xi_4}
    \tilde{\mathcal{T}}^{(\alpha)}_{(\bar\xi_3 i_3)(\xi_1\xi_2\bar\xi_4i_1i_2i_4mn)}\tilde{\mathcal{T}}^{(\alpha)\dagger}_{(\bar\xi_1\bar\xi_2\xi_4i_1i_2i_4mn)(\xi_3' i_3')} \ ,\\
    (\tilde{\mathcal{M}}_{+y})_{(\bar\xi_2 i_2)(\xi_2' i_2')}&=\!\!\!\!
    \sum_{i_1,i_3,i_4,m,n}\int_{\bar\xi_1\xi_1,\bar\xi_3\xi_3,\bar\xi_4\xi_4}
    \tilde{\mathcal{T}}^{(\alpha)\dagger}_{(\bar\xi_2 i_2)(\bar\xi_1\xi_3\xi_4i_1i_3i_4mn)}\tilde{\mathcal{T}}^{(\alpha)}_{(\xi_1\bar\xi_3\bar\xi_4i_1i_3i_4mn)(\xi_2' i_2')}  \ ,\\
    (\tilde{\mathcal{M}}_{-y})_{(\bar\xi_4 i_4)(\xi_4' i_4')}&=\!\!\!\!
    \sum_{i_1,i_2,i_3,m,n}\int_{\bar\xi_1\xi_1,\bar\xi_2\xi_2,\bar\xi_3\xi_3}
    \tilde{\mathcal{T}}^{(\alpha)}_{(\bar\xi_4 i_4)(\xi_1\xi_2\bar\xi_3i_1i_2i_3mn)}\tilde{\mathcal{T}}^{(\alpha)\dagger}_{(\bar\xi_1\bar\xi_2\xi_3i_1i_2i_3mn)(\xi_4' i_4')} 
    \  .
    \label{eq:Mtildelast}
\end{align}
By diagonalizing these matrices, we obtain the hybrid
isometries $(\mathcal{U}_{\pm \mu})_{\bar\xi i;\phi}$ that further
combine the bosonic and fermionic indices together.
Just like $(\mathcal{V}_{\pm a})_{\bar\zeta j;\xi}$,
the combined index $\phi$ is also truncated with the bond dimension $\chi_c$.
Thus the final compressed tensor
becomes (See Fig.~\ref{fig:compactification}-c and -d.)
\begin{align}
    \mathcal{T}^{(\alpha)'}_{\phi_1\phi_2\bar\phi_3\bar\phi_4;mn}&=\sum_{IJKL}T^{(\alpha)'}_{IJKL;mn}\phi_1^I\phi_2^J\bar\phi_3^K\bar\phi_4^L
    \label{eq:compressdT}
    \\
    &\!\!\!\!\!\!\!\!\!\!\!\!\!\!\!\!\!\!\!\!\!\!\!\!\!\!\!\!\!\!=
    \int_{\{\bar\xi_\nu\xi_\nu\}}
    \sum_{\{i_\nu\}}
    (\mathcal{U}_y)^\dagger_{(\bar\phi_4)(\xi_4i_4)}
    (\mathcal{U}_x)^\dagger_{(\bar\phi_3)(\xi_3i_3)}
    \tilde{\mathcal{T}}^{(\alpha)}_{\xi_1\xi_2\bar\xi_3\bar\xi_4;i_1i_2i_3i_4;mn}
    (\mathcal{U}_x)_{(\bar\xi_1i_1)(\phi_1)}
    (\mathcal{U}_y)_{(\bar\xi_2i_2)(\phi_2)} \ .
\end{align}
Here $\zeta$'s are the four legs pointing
in the 2D space, while $m$ and $n$ are the bosonic legs
joining different flavor layers.

\subsection{Some comments on the computational cost}

In this subsection, we make some comments on the computational cost.
The most computationally expensive parts of the process are
the contraction of $\mathcal{M}_{\pm a}$ \eqref{eq:Mfirst}-\eqref{eq:Mlast}
and $\tilde{\mathcal{M}}_{\pm a}$ \eqref{eq:Mtildefirst}-\eqref{eq:Mtildelast},
which contain 11 and 12 loops, respectively.
However, the number of loops in the computation of $\mathcal{M}_{\pm a}$
can be reduced
since the index $m$ is always the same as one of the other indices
due to the Kronecker delta, which can effectively reduce the depth to 10 loops.
In terms of complexity, the cost of the computation
of $\mathcal{M}_{\pm a}$ and $\tilde{\mathcal{M}}_{\pm a}$
are $16^5K^5$ and $D_\xi^5K^7$, respectively,
where $D_\xi$ is the bond dimension of the fermionic legs $\xi$.

At the technical level, we can do a few more things
to further reduce the computational cost.
First, we implement another compression
on the
fermionic legs of the original $\mathcal{S}$ tensor \eqref{eq:Btensor}
to reduce the bond dimension of the leg $\zeta_a$, which is
of dimension $D_\zeta=16$.
In addition, we improve
the speed of compression significantly
by storing the tensors as sparse arrays
and performing contractions with a sparse matrix-based
algorithm \cite{10.1145/3571157}.

\subsection{The coarse-graining procedure}
\label{sec:coarse-graining}

\begin{figure}
    \centering
    \includegraphics[scale=0.8]{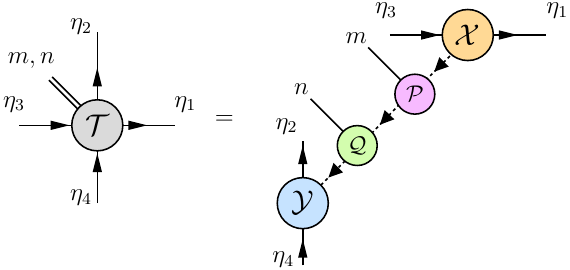}
    \caption{The decomposition of the site tensor in the modified HOTRG.}
    \label{fig:gHOTRG_simplified}
\end{figure}

In order to perform the coarse-graining procedure in the flavor direction,
we implement a modified version of HOTRG.
Note first that
the site tensor $\mathcal{T}'_{\eta_1\eta_2\bar\eta_3\bar\eta_4;mn}$ has six indices.
Four of them are the fermionic indices in the 2D space directions,
while $m$ and $n$ are the indices
for the gauge link variables.
Since the site tensors in the flavor direction are connected
with the Kronecker deltas,
the bosonic bond degrees of freedom are maximally entangled in the flavor direction,
which implies that we cannot insert isometries to
compress these bonds further.
For this reason, we have to
perform the coarse-graining procedure in the flavor direction
before the coarse-graining in the two-dimensional space.

Since the size of the compressed initial tensor \eqref{eq:compressdT}
grows like $K^2$ with $K$ being the bond dimension of the bosonic legs,
the coarse-graining can become costly with the traditional HOTRG
algorithm \cite{PhysRevB.86.045139} as we increase $K$.
(See Appendix \ref{section:algorithm} for a review of the HOTRG and
related methods.)
We therefore modify the HOTRG method by
first performing the decomposition such that we separate the legs into
three groups based on their axes as
\begin{align}
    \mathcal{T}^{(\alpha)}_{\eta_1\eta_2\bar\eta_3\bar\eta_4;mn}=\int_{\bar\zeta\zeta,\bar\phi\phi,\bar\xi\xi}
    \mathcal{X}^{(\alpha)}_{\eta_1\bar\eta_3\zeta}
    \mathcal{P}^{(\alpha)}_{\bar\zeta\phi;m}
    \mathcal{Q}^{(\alpha)}_{\bar\phi\xi;n}
    \mathcal{Y}^{(\alpha)}_{\bar\xi\eta_2\bar\eta_4} \ ,
\end{align}
where the legs along the $x$ axis are in the $\mathcal{X}$ tensor,
the legs along the $y$ axis are in the $\mathcal{Y}$ tensor,
and the legs along the flavor axis are in the $\mathcal{P}$
and $\mathcal{Q}$ tensors.
With this decomposition, the isometries along the $x$ and $y$ axis are
computed using the tensor $\mathcal{X}$ and $\mathcal{Y}$, respectively.


\begin{figure}
    \centering
    \includegraphics[scale=0.7]{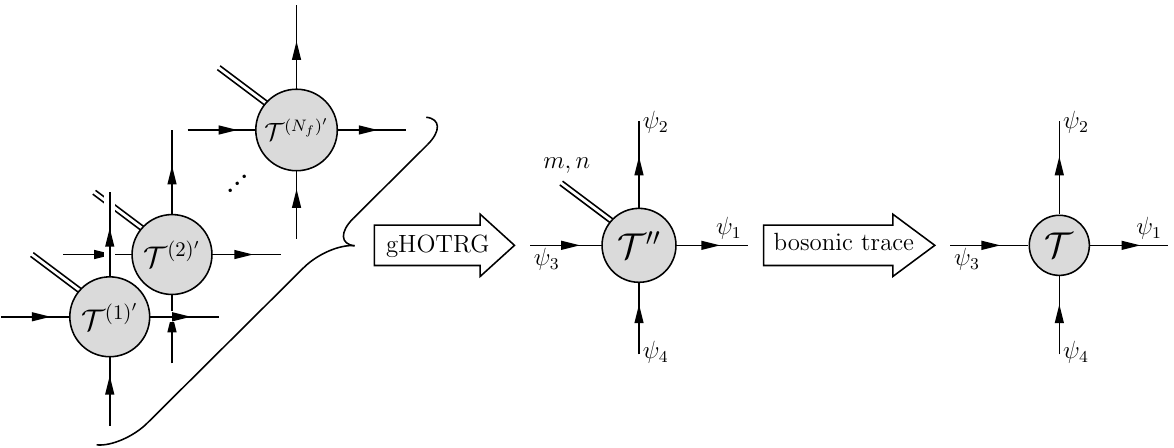}
    \caption{The summary of the flavor coarse-graining procedure.}
    \label{fig:zblocking}
\end{figure}

In order to obtain the partition function,
we first perform coarse-graining procedure in the $z$ (flavor) direction
using the Grassmann higher-order TRG (gHOTRG) algorithm
described in Appendix \ref{section:HOTRG},
which gives us the $N_{\rm f}$-flavor
tensor $\mathcal{T}''_{\psi_1\psi_2\bar\psi_3\bar\psi_4;mn}$.
Then we take the trace of the bosonic indices
\begin{equation}
  \mathcal{T}_{\psi_1\psi_2\bar\psi_3\bar\psi_4}
  =\sum_{m,n}\mathcal{T}''_{\psi_1\psi_2\bar\psi_3\bar\psi_4;mn} \ ,
  \label{eq:zblocking}
\end{equation}
which is described schematically in Fig.~\ref{fig:zblocking}.

Next, we perform the coarse-graining in the 2D plane
using the Grassmann TRG (gTRG)
which is
described in detail in Appendix \ref{section:2dTRG}.
The bond dimensions for the gHOTRG in the flavor direction is $\chi_f$,
while the bond dimension for gTRG in the 2D plane is $\chi_{xy}$.
After performing the coarse-graining procedure sufficiently many times,
we take the trace of the final tensor with anti-periodic boundary conditions
in the imaginary time direction to obtain the partition function as
\begin{equation}
    Z = \sum_{I,J}T_{IJIJ}\sigma_I\sigma_J(-)^{p(I)p(J)+p(J)} \ ,
\end{equation}
where $\sigma_I$ is the sign factor defined in \eqref{eq:sgn_defninition}.

All of the computations in this paper are done using a
Python package \verb|grassmanntn| \cite{grassmannTN}
specialized in handling the Grassmann tensor network.

\section{Numerical results}
\label{section:results}

In this section, we present our numerical results obtained by the
method introduced in the previous sections.
First, we perform performance tests
concerning
the initial tensor compression
described in section \ref{section:compression}
and
the coarse-graining procedure in the flavor direction
described in section \ref{sec:coarse-graining}.
Then we demonstrate the usefulness of our method by
investigating
the chiral phase transition and the Silver Blaze phenomenon
in 2D Abelian gauge theories.
In what follows, we assume that all the flavors of fermions have the same charge $q_\alpha=q$,
mass $\tilde m_\alpha=\tilde m$ and
chemical potential $\tilde\mu_\alpha=\tilde\mu$.

\subsection{Performance tests}

Let us first
demonstrate the efficiency of the initial tensor compression.
For that, we compute $\log Z$ for various parameters and measure the errors by
comparing the results
obtained with and without compression.
It is found that the relative error of the compression is less than $10^{-15}$ for all the cases if we choose the compression bond dimension $\chi_c=64$.
The efficiency of the compression for this choice
is summarized in Table \ref{tab:compression}.
It is clear that our compression scheme is efficient, in particular for large $K$,
where the original tensor can easily become too large to be handled by
the currently available computers.

\begin{table}
    \centering
    \begin{tabular}{|cccc||c|c|c|c|c|}
        \hline
        $\beta$ & $\tilde\mu$ & $N_{\rm f}$ & $K$ & original size& compressed size& compression ratio&$D_x$&$D_y$\\
        \hline
        0.0&0.0&1&2&67108864      & 1024   & $1.53\times10^{-5}$ &4 &4\\
        0.0&0.0&1&3&3869835264   & 2304  & $5.95\times10^{-7}$ &4 &4\\
        0.0&0.0&1&4&68719476736   & 4096  & $5.96\times10^{-8}$ &4 &4\\
        0.0&0.0&1&5&640000000000 & 6400 & $1.00\times10^{-9}$ &4 &4\\
        
        2.0&0.0&1&2&67108864      & 16384  & $2.44\times10^{-4}$ &8 &8\\
        2.0&0.0&2&2&67108864      & 16384  & $2.44\times10^{-4}$ &8 &8\\

        2.0&3.0&1&2&67108864      & 16384  & $2.44\times10^{-4}$ &8 &8\\
        2.0&3.0&2&2&67108864      & 16384  & $2.44\times10^{-4}$ &8 &8\\
        \hline
    \end{tabular}
    \caption{Summary of the initial tensor compression for
      various input parameters $\beta$, $\tilde\mu$, $N_{\rm f}$, and $K$. The relative error of the compression is less than $10^{-15}$ for all the cases.
      The size of the original tensor is obtained by the formula $16^{4}K^{10}$,
      whereas that of the compressed tensor is obtained by $(D_xD_yK)^2$.
      Here $D_x$ and $D_y$
      represent the bond dimension of the legs $I_1$ and $I_3$
      and the bond dimension of the legs $I_2$ and $I_4$, respectively,
      of the compressed coefficient tensor $T^{(\alpha)'}_{I_1I_2I_3I_4;mn}$, while
      $K$ represents the bond dimension of the bosonic legs $m$ and $n$.}
    \label{tab:compression}
\end{table}


\begin{figure}
    \centering
    \includegraphics[scale=0.8]{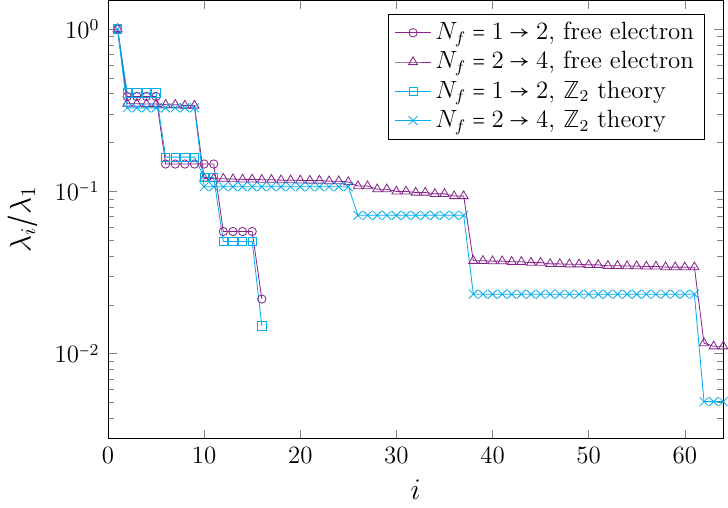}
    \caption{The singular value spectrum associated with the HOTRG isometry
      truncation in the free electron gas model
      and the $\mathbb{Z}_2$ gauge theory.
      Here we show only the spectra of the $x$-axis truncation,
      which is identical to that of the $y$-axis truncation.}
    \label{fig:compare_Nf}
\end{figure}

Next, we discuss
the performance of coarse-graining in the flavor direction.
Here we perform the gHOTRG up to $N_{\rm f}=4$ for the free electron gas model ($K=1$)
and the $\mathbb{Z}_2$ gauge theory with $\beta=0$, $q=1$, $\tilde m=1$
and $\tilde\mu=0$.
In Fig.~\ref{fig:compare_Nf},
we plot the singular value spectra associated with the SVD
when the isometries ($\mathcal{U}_x$ and $\mathcal{U}_y$
in \eqref{eq:Xtilde} and \eqref{eq:Ytilde})
are used
during the step $N_{\rm f}=1\rightarrow2$ and $2\rightarrow4$ with $\chi_f=64$.
One can see that the tail of the singular value spectrum grows quickly
with $N_{\rm f}$, which indicates that fermions from different layers have
strong degeneracy. Note that introducing gauge interaction makes the singular
value spectrum decays faster.
For the calculations in the subsequent subsections,
we use $\chi_f=64$ for $N_{\rm f}=2$ and $\chi_f=32$ for $N_{\rm f}=4$
for the flavor coarse-graining,
and $\chi_{xy}=64$ for the two-dimensional coarse-graining.

\subsection{The chiral phase transition}

\begin{figure}
    \centering
    \includegraphics[scale=0.64]{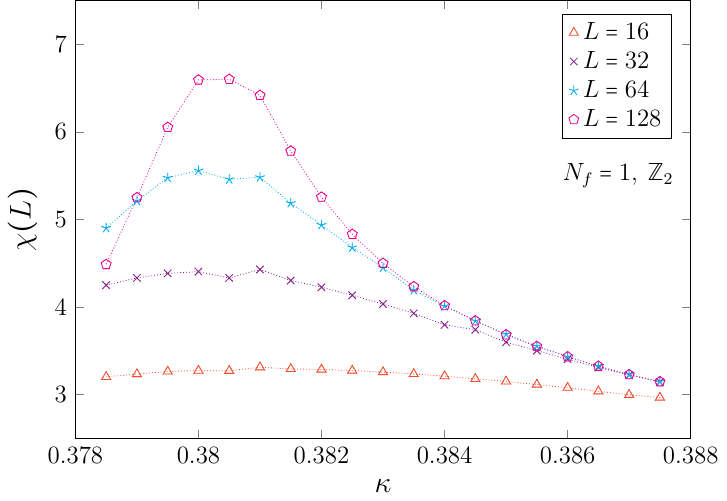}
    \includegraphics[scale=0.64]{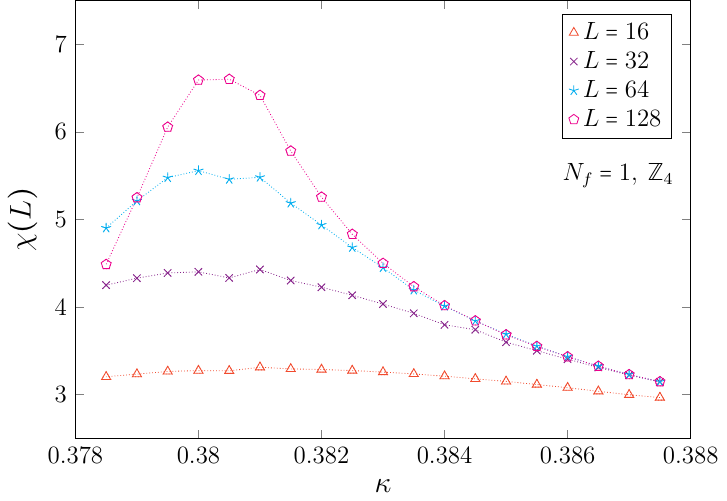}\\
    \includegraphics[scale=0.64]{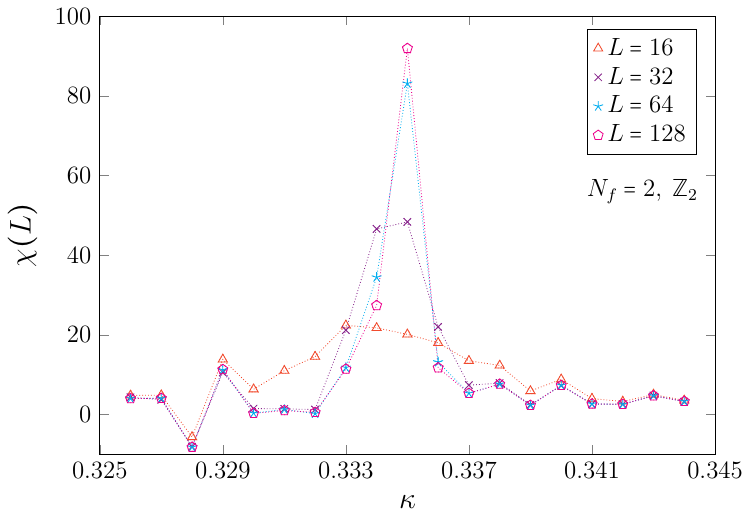}
    \includegraphics[scale=0.64]{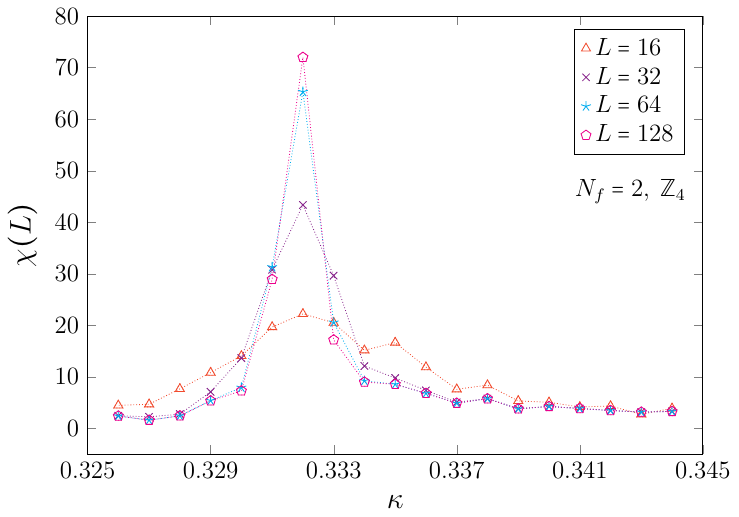}\\
    \includegraphics[scale=0.64]{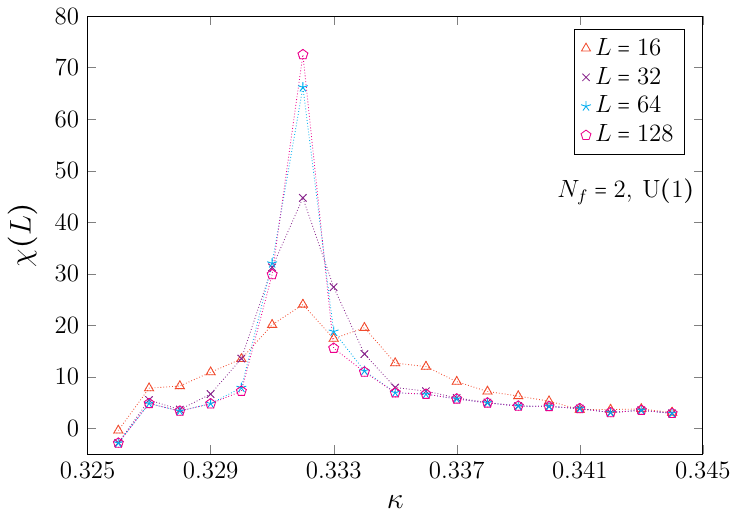}
    \caption{Chiral susceptibility is plotted
      as a function of $\kappa$ at $\beta=0$ for
      different gauge theories and $N_{\text{f}}$ (labeled in each plot) with lattice volume
      up to $V=128^2$. The dotted lines are shown to guide the eyes. The critical hopping parameter $\kappa_\text{c}$ can be identified as the location of the peak in the infinite volume limit. Note that the plots for $N_\text{f}=1$ are almost identical for the $\mathbb{Z}_2$ and $\mathbb{Z}_4$ cases.
      }
    \label{fig:finding_critical_kappa}
\end{figure}


In order to demonstrate the usefulness of our method,
we first apply it to
the chiral phase transition in
two-flavor $\mathbb{Z}_2$, $\mathbb{Z}_4$, and $\text{U}(1)$ gauge theories.
Let us define the hopping parameter by
\begin{equation}
    \kappa=\frac{1}{2\tilde m+4}  \ .
\end{equation}
Although the chiral symmetry is broken explicitly
by the Wilson term in \eqref{fermion-action},
it is expected to be restored
at some critical hopping parameter $\kappa_\text{c}$.
We can easily identify $\kappa_\text{c}$
in the TRG method
by the location of the peak of chiral susceptibility
\begin{equation}
    \chi(L)=\frac{1}{V}\frac{\partial^2}{\partial\tilde m^2}\log Z
\end{equation}
given as a function of the hopping parameter $\kappa$,
where the derivatives can be taken numerically.
The peak of the chiral susceptibility
exhibits a critical behavior in the large volume limit as
demonstrated in Ref.~\cite{Shimizu:2014uva} with $N_{\rm f}=1$.
In Fig.~\ref{fig:finding_critical_kappa}.
we observe similar behavior in various gauge theories with $N_\text{f}=1$ and 2
at $\beta=0$, where we have used
$\chi_f=\chi_{xy}=64$. For the $\text{U}(1)$ case, we use the 4-nodes Gauss-Legendre quadrature to discretize the group integral.

In the large-$K$ limit, $\mathbb{Z}_K$ gauge theory converges to the $\text{U}(1)$ gauge theory. For $N_{\rm f}=1$, the
critical hopping parameter $\kappa_\text{c}=0.3806(1)$
obtained by our method in both $\mathbb{Z}_2$ and $\mathbb{Z}_4$ theories
is consistent with the result $\kappa_\text{c}=0.380665(59)$
obtained for the ${\rm U}(1)\simeq\mathbb{Z}_\infty$ theory in Ref.~\cite{Shimizu:2014uva}, which indicates that
the convergence occurs already
at $K=2$ for $\beta=0$.
For $N_{\rm f}=2$, we obtain
$\kappa_\text{c}$ for the $\mathbb{Z}_2$, $\mathbb{Z}_4$, and ${\rm U}(1)$
theories at $\beta=0$
and compare with the Monte Carlo results for the ${\rm U}(1)$
theory \cite{Hip:1997em}
in table \ref{tab:critical_hopping}.


\begin{table}
    \centering
    \begin{tabular}{|c|c|c|}
    \hline
    Gauge group & Algorithm & $\kappa_\text{c}$\\
    \hline
    $\mathbb{Z}_2$ & gTRG & 0.335(1)\\
    $\mathbb{Z}_4$ & gTRG & 0.332(1)\\
    ${\rm U}(1)$ & gTRG & 0.332(1)\\
    ${\rm U}(1)$ & Monte Carlo \cite{Hip:1997em} & 0.3296983759\\
    \hline
    \end{tabular}
    \caption{The critical hopping parameter
    for various gauge theories with $N_{\rm f}=2$ at $\beta=0$. Our $\text{U}(1)$ result is computed with the 4-nodes Gauss-Legendre quadrature.}
    \label{tab:critical_hopping}
\end{table}


\subsection{Silver Blaze phenomenon at finite density}

Next we apply our method to the finite density case with multiple flavors.
In particular,
it is expected that
physical observables
in the thermodynamic limit
and at zero temperature are independent of the chemical potential
up to some threshold due to the gapped spectrum in the confined phase.
This is known as the Silver Blaze phenomenon \cite{Cohen:2003kd},
which is difficult to reproduce by Monte Carlo methods
due to the sign problem.
We investigate this phenomenon by our method
to demonstrate that the sign problem is indeed solved.

\begin{figure}
    \centering
    \includegraphics[scale=0.65]{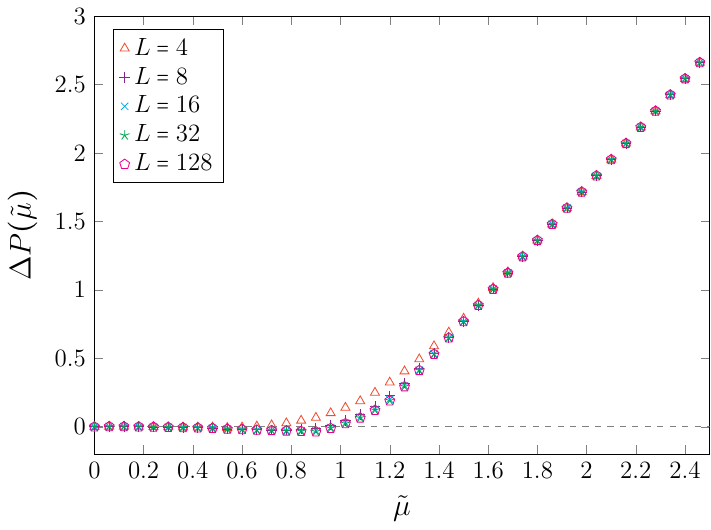}
    \includegraphics[scale=0.65]{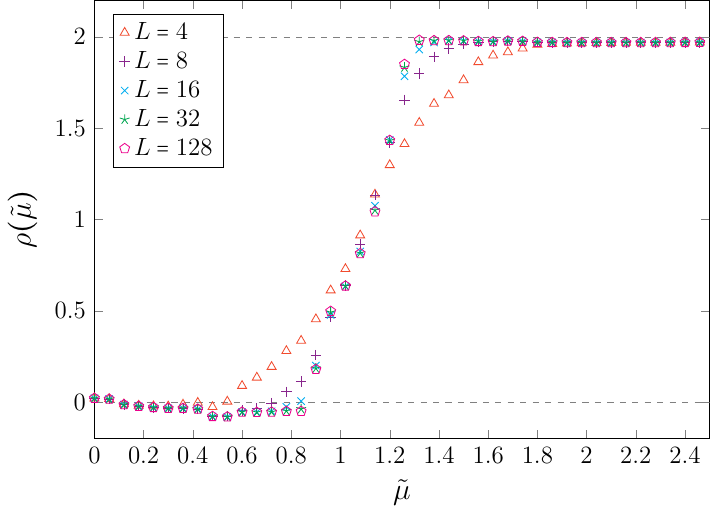}
    \caption{Differential pressure $\Delta P(\tilde\mu)$ and
      number density $\rho(\tilde\mu)$ of the $\mathbb{Z}_2$ gauge theory with $N_{\rm f}=2$ at
      various volume $V=L\times L$. The small non-monotonicity
      may be attributed to the finite truncation effect.}
    \label{fig:finite_density_N2}
\end{figure}

\begin{figure}
    \centering
    \includegraphics[scale=0.65]{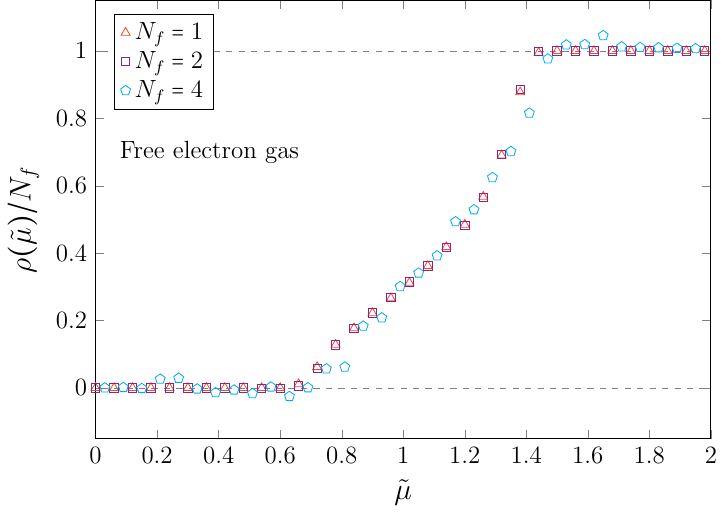}
    \includegraphics[scale=0.65]{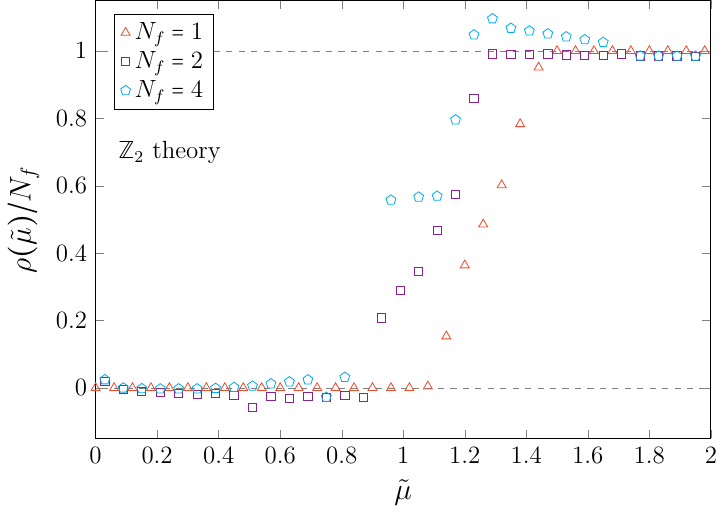}
    \caption{$\rho(\tilde\mu)/N_{\rm f}$
      is plotted against
      $\tilde\mu$ for $V=128\times 128$
in the free electron model and
the $\mathbb{Z}_2$ gauge theory with $N_{\rm f}=1$, $2$, and $4$. The non-monotonicity may be attributed to the truncation effect, which becomes more significant for larger $N_{\rm f}$.}
    \label{fig:number_density_various_models}
\end{figure}

Here 
we consider the $\mathbb{Z}_2$ gauge theory and
the free electron model
with $\beta=1$, $q=1$, $\tilde m=1$.
We calculate the pressure and the number density
\begin{align}
  P(\tilde\mu) &=\frac{1}{V}\log Z \ ,
  \label{eq:pressure}\\
    \rho(\tilde\mu) &=\frac{\partial P}{\partial\tilde\mu} \ ,
    \label{eq:number_density}
\end{align}
where the derivative in \eqref{eq:number_density}
is taken numerically.

In Fig.~\ref{fig:finite_density_N2}
we plot
the differential
pressure $\Delta P(\tilde\mu)=P(\tilde\mu)-P(0)$ and
the number density $\rho(\tilde\mu)$
against $\tilde\mu$ up to the volume $V=128\times128$
in the $\mathbb{Z}_2$ gauge theory with $N_{\rm f}=2$.
We observe a clear Silver Blaze phenomenon for $\tilde\mu\lesssim0.8$.
We also find that the number density
saturates
to the value $\rho=2$
for $\tilde\mu>1.4$ as expected from the number of degrees of freedom
at each lattice site.

In Fig.~\ref{fig:number_density_various_models}
we plot $\rho(\tilde\mu)/N_{\rm f}$ against
$\tilde\mu$
in the free electron model and
the $\mathbb{Z}_2$ gauge theory with $N_{\rm f}=1$, $2$, and $4$.
The number density saturates to the value $\rho=N_{\rm f}$
for all cases.
Note also that in the free electron gas model,
$\rho(\tilde\mu)/N_{\rm f}$ are expected to be the same for all $N_{\rm f}$,
which is not the case
in the $\mathbb{Z}_2$ gauge theory due to interactions.

\section{Summary}
\label{section:summary}

In this paper, we have proposed a new technique to
incorporate multiple flavors
in the TRG method for lattice gauge theories.
The problem of the initial tensor, which grows
in size exponentially with the number of flavors $N_{\rm f}$,
has been overcome
by separating the initial tensor into $N_{\rm f}$ layers
with replicas of the gauge field for each layer, which are identified later.
This effectively makes the system one dimension higher
due to the flavor direction.
Consequently,
the tensor can still be large, in particular,
due to the gauge field legs in the extra dimension.
In order to overcome this problem,
we
use a compression scheme,
which proceeds in two steps
by truncating first the subtensors and then the whole tensor.
We have shown that
this enables us to compress the size of the
initial tensor by many orders of magnitude without sacrificing the accuracy.
Notably, the compression is found to be more effective for larger $K$
in $\mathbb{Z}_K$ gauge theories,
where the original tensor can be too large to perform any calculation
with currently available computers.

As another important performance test,
we have studied
the singular value spectrum of the flavor coarse-graining procedure
and find that introducing gauge interaction makes
the spectrum decays faster.
In order to demonstrate the usefulness of our method,
we have investigated the chiral phase transition
in two-dimensional Abelian gauge theories with $N_{\rm f}=2$
by computing the critical value of the hopping parameter,
which turns out to be consistent with the known value obtained by the Monte Carlo method.
We have also applied our method to the case of finite density
with $N_{\rm f}=1$, $2$, and $4$
in the $\mathbb{Z}_2$ gauge theory.
In particular,
we were able to observe the Silver Blaze phenomenon,
which is difficult to reproduce
by Monte Carlo methods due to the sign problem.


We consider that our new technique
will make the TRG method applicable to many interesting gauge theories
with multiple flavors that have not been explored yet.
Since the main idea can be generalized to the non-Abelian case,
we hope that it will be useful
also in investigating
QCD, where two (or three) flavors of light quarks have to be incorporated.
We also expect that, by implementing better renormalization schemes such as the bond-weighting methods \cite{Adachi:2020upk, Akiyama:2022pse} or stochastic sampling approach \cite{ferris2015unbiased,huggins2017monte,Arai:2022uee}, the systematic error from bond truncation can be reduced.
Last but not the least,
we expect that our technique
is useful in applying the TRG method
to the domain-wall formalism for chiral fermions,
where the extra dimension can be regarded as the flavor direction in our method.
In that case, we need to introduce local interactions
in the flavor direction as described in
Appendix \ref{section:interflavor_interaction}.
If such interactions
make the singular-value spectrum in the flavor direction
decay faster,
one can go to larger $N_{\rm f}$, which is crucial
in the domain-wall formalism.
We hope to report on this in the future publication.


\section*{Acknowledgments}
We would like to thank Akira Matsumoto for valuable discussions and Abhabongse Janthong for valuable advice in code development and optimization.
A.~Y. and K.~O. are supported by a Grant-in-Aid for Transformative Research Areas
``The Natural Laws of Extreme Universe—A New Paradigm for Spacetime and Matter from
Quantum Information” (KAKENHI Grant No. JP21H05191) from JSPS of Japan.

\appendix

\bigskip

\section{Grassmann tensor network}
\label{section:grassmann_review}
In this section, we provide a
a formulation of
the Grassmann tensor network,
where we explore the Grassmann tensor and its properties in more detail
than in
the original paper \cite{Akiyama:2020sfo}.
In particular, we propose a new format for the coefficient tensor
that has an intuitive connection with the non-Grassmann linear algebra.

In general, one can expand any function of Grassmann variables
as a polynomial
\begin{equation}
    \mathcal{T}(\theta_1,\theta_2,\cdots,\theta_n)\equiv\mathcal{T}_{\theta_1\theta_2\cdots\theta_n}=\sum_{i_1i_2\cdots i_n\in\{0,1\}}T_{i_1i_2\cdots i_n}\theta_1^{i_1}\theta_2^{i_2}\cdots\theta_n^{i_n} \ ,
    \label{eq:Grassmann_tensor_expansion_original}
\end{equation}
where the complex-valued coefficient $T_{i_1i_2\cdots i_n}$
shall be referred to as the \emph{coefficient tensor}.
Multiple Grassmann numbers can be grouped into a single multi-component
variable, for instance, as
\begin{equation}
    \psi^{I}\equiv\theta_{1}^{i_1}\cdots \theta_{m}^{i_m} \ ,
\end{equation}
where $I=(i_1,\cdots i_m)$.
Hereafter,
we reserve the symbol $\theta$ for a one-component Grassmann number
and use
other Greek letters
for multi-component Grassmann numbers.
In practice, the composite index $I$ must be encoded as an integer
with some binary encoder $f(i_1,\cdots i_m)\in\mathbb{Z}$.
Physical quantities are independent of the encoding function,
but the calculation can be made easier if it is chosen appropriately.
Two useful ones are the ``canonical'' and
the ``parity-preserving'' \cite{Akiyama:2020sfo} encoders defined,
respectively, by
\begin{align}
    f_\text{canonical}(i_1,\cdots,i_n)&=\sum_{k=1}^n2^{k-1}i_k \ ,
    \label{eq:canonical_encoder}
    \\
    f_\text{parity-preserving}(i_1,\cdots,i_n)&=
    \left\{
    \begin{array}{ll}
    \displaystyle
    \sum_{k=1}^n2^{k-1}i_k&
    \displaystyle
    ;i_2+\cdots +i_n\;\text{even} \ ,\\
    \displaystyle
    1-i_1+\sum_{k=2}^n2^{k-1}i_k&
    \displaystyle
    ;i_2+\cdots +i_n\;\text{odd} \ .
    \end{array}
    \right.
    \label{eq:parity_preserving_encoder}
\end{align}
The canonical encoder is more intuitive and easier to join and split the indices,
while
the parity-preserving encoder is essential in the
construction of the isometries, which is explained below.
In this paper, both the composite index and its encoding
are
referred to by the same capital Latin letters
for
simplicity.

In order to
combine two indices that are not adjacent,
we have to rearrange these indices first so that the indices to be combined
are next to each other.
When the two indices are swapped, the coefficient tensor must be multiplied
by a sign factor due to the anti-commuting nature of the Grassmann numbers.
For example, if we swap $T_{IJKL}$ into $T'_{IKJL}$, we have to multiply the
relative sign factor coming from
$\psi_1^J\psi_2^K=(-)^{p(J)p(K)}\psi_2^K\psi_1^J$ to the coefficient tensor as
\begin{equation}
    T_{IJKL}=T'_{IKJL}(-)^{p(J)p(K)} \ ,
\end{equation}
where $p(I)=\sum_ai_a$ is the Grassmann parity of $\psi^I$.

\subsection{Grassmann tensors and their contraction}
The contraction of the Grassmann indices is performed
by the Berezin integral of the pair $(\bar\theta,\theta)$.
For example, in the case of one component, we have the identity
\begin{equation}
  \int d\bar\theta d\theta e^{-\bar\theta\theta}\theta^i\bar\theta^j
  = \delta_{ij} \ .
    \label{eq:one-component_contraction}
\end{equation}
Then the contraction of the two Grassmann tensors reads
\begin{equation}
    \int d\bar\theta d\theta e^{-\bar\theta\theta}\mathcal{A}_{\bar\theta_1\theta}\mathcal{B}_{\bar\theta\theta_2}=\sum_{i,j}(\sum_kA_{ik}B_{kj})\bar\theta_1^i\theta_2^j \ .
\end{equation}

The multi-component case is slightly more complicate since, for $I=(i_1,\cdots,i_n)$ and $\psi^I=\theta_1^{i_1}\cdots\theta_n^{i_n}$, the contraction
\begin{equation}
    \prod_{a=1}^n\int d\bar{\theta}_a d\theta_a e^{-\bar{\theta}_a\theta_a}
    \psi^{I}\bar{\psi}^{J}=\left(\prod_a\delta_{i_a,j_a}\right)\times\left(\prod_{a<b}(-)^{i_ai_b}\right) \ ,
    \label{eq:multi-index_contraction_full}
\end{equation}
has the extra sign factor coming from rearranging $\theta_a$ and $\bar\theta_a$
for the integration.
Defining
\begin{align}
    \int d\bar{\psi} d\psi e^{-\bar{\psi}\psi}&\equiv\prod_{a=1}^n\int d\bar{\theta}_a d\theta_a e^{-\bar{\theta}_a\theta_a} \ , \\
    \delta_{IJ}&\equiv\prod_a\delta_{i_a,j_a} \ ,\\
    \sigma_{I}&\equiv\prod_{a<b}(-)^{i_ai_b} \ ,
    \label{eq:sgn_defninition}
\end{align}
the contraction \eqref{eq:multi-index_contraction_full} can be rewritten
in a compact form
\begin{equation}
  \int d\bar\psi d\psi
  e^{-\bar\psi\psi}\psi^I\bar\psi^J=\delta_{IJ}\sigma_I \ ,
    \label{eq:multi-index_contraction}
\end{equation}
which is the multi-component
counterpart
of \eqref{eq:one-component_contraction}.
The contraction rule according to the identity \eqref{eq:multi-index_contraction}
is
\begin{equation}
    \int d\bar\eta d\eta e^{-\bar\eta\eta}\mathcal{A}_{\bar\psi\eta}\mathcal{B}_{\bar\eta\phi}=\sum_{I,J}(\sum_KA_{IK}B_{KJ}\sigma_K)\bar\psi^I\phi^J \ ,
    \label{eq:original_matrix_contraction}
\end{equation}
which differs slightly from the usual matrix contraction
due to the extra sign factor $\sigma_K$.

It is possible to define a new format for the coefficient tensor
such that the contraction can be done without the extra sign factor as
\begin{equation}
    \mathcal{A}_{\bar\psi_1\cdots\bar\psi_m\phi_1\cdots\phi_n}=\sum_{I_1\cdots I_mJ_1\cdots J_n}A^\text{(m)}_{I_1\cdots I_mJ_1\cdots J_n}\sigma_{I_1}\cdots\sigma_{I_m}\bar\psi_1^{I_1}\cdots\bar\psi_m^{I_m}\phi_1^{J_1}\cdots\phi_n^{J_n} \ ,
\end{equation}
which we call the \emph{matrix format}.
The coefficient tensor in this format
can be defined in terms of that in the standard format as
\begin{equation}
    A^\text{(m)}_{I_1\cdots I_mJ_1\cdots J_n}\equiv A_{I_1\cdots I_mJ_1\cdots J_n}\sigma_{I_1}\cdots\sigma_{I_m} \ .
    \label{eq:def_grassmann_matrix}
\end{equation}
Namely
we multiply the sign factor $\sigma_{I_a}$ for every conjugated
Grassmann index $\bar\psi_a^{I_a}$.
It is straightforward to check that the coefficient matrix of
the contraction \eqref{eq:original_matrix_contraction} is
actually the matrix product $A^\text{(m)}B^\text{(m)}$ as
\begin{equation}
    \mathcal{C}_{\bar\psi\phi}=\int d\bar\eta d\eta e^{-\bar\eta\eta}\mathcal{A}_{\bar\psi\eta}\mathcal{B}_{\bar\eta\phi}=\sum_{I,J}
    (A^{(\text{m})}B^{(\text{m})})_{IJ}
    \sigma_I\bar\psi^I\phi^J=\sum_{I,J}
    C^{(\text{m})}_{IJ}
    \sigma_I\bar\psi^I\phi^J \ .
\end{equation}
This simple contraction rule also applies to tensors of arbitrary rank
if it is written in the matrix format. Note that the sign factors
from the index permutation must also be applied
if the contracted indices are not adjacent.

Tensor legs can be joined together with the prescription
\begin{align}
    \xi^K&\equiv\psi_1^{I_1}\cdots\psi_m^{I_m}\bar\phi_1^{J_1}\cdots\bar\phi_n^{J_n} \ ,
    \label{eq:join_leg1}\\
    \bar\xi^K&\equiv\bar\psi_1^{I_1}\cdots\bar\psi_m^{I_m}\phi_1^{J_1}\cdots\phi_n^{J_n}\prod_{a=1}^n(-)^{p(J_a)} \ ,
    \label{eq:join_leg2}\\
    \int d\bar\xi d\xi e^{-\bar\xi\xi}&\equiv\int \prod_{a=1}^m\left(d\bar\psi_a d\psi_a e^{-\bar\psi_a\psi_a}\right)\prod_{b=1}^n\left(d\bar\phi_b d\phi_b e^{-\bar\phi_b\phi_b}\right) .
    \label{eq:join_leg3}
\end{align}
It is important that the sign factor $(-)^{p(J_a)}$,
where $J_a$ is the index of the non-conjugated constituent $\phi_a$,
must be introduced if the joined variable is a conjugated fermion.
This is to ensure that the contraction between $\xi$ and $\bar\xi$ with the measure \eqref{eq:join_leg3}
follows the contraction rule \eqref{eq:original_matrix_contraction}.

Throughout this paper,
we abbreviate the contraction integral by
\begin{equation}
    \int_{\bar\eta\eta}\equiv\int d\bar\eta d\eta e^{-\bar\eta\eta} \ .
    \label{eq:contraction_definition}
\end{equation}
Note that the Grassmann contraction is directional
since the Grassmann variables are anti-commuting. Therefore
the bond between
any two Grassmann tensors should have an arrow pointing from $\eta$ to $\bar\eta$.

\subsection{Eigenvalue and singular-value decompositions}
\label{section:eigensystem}

Here we introduce
the concept of matrix decomposition for Grassmann tensors,
which is similar to the one used in the traditional tensor network
as an important part
of the coarse-graining procedure.
In this context, we assume that the tensor has already been
reshaped (with \eqref{eq:join_leg1}-\eqref{eq:join_leg2}) into a matrix
\begin{equation}
    \mathcal{M}_{\bar\psi\phi}=\sum_{IJ}M^\text{(m)}_{IJ}\sigma_I\bar\psi^I\phi^J \ .
\end{equation}
Let $v^{\text{(m)}\dagger}$ and $u^\text{(m)}$ be the
left and right eigenvectors of $M^{(\text{m})}$ with the eigenvalue $\lambda$ as
\begin{equation}
  v^{\text{(m)}\dagger} M^{(\text{m})}=\lambda v^{\text{(m)}\dagger} \ ,
  \qquad M^{(\text{m})}u^\text{(m)}=\lambda u^\text{(m)} \ .
\end{equation}
The Grassmann eigenvectors of $\mathcal{M}$ can then be defined by
\begin{equation}
  v^\dagger_{{\psi}}=\sum_{I}v^{\text{(m)}*}_{I}{\psi}^{I} \ ,
  \qquad u_{\bar{\psi}}=\sum_{I}u^\text{(m)}_{I}\sigma_{I}\bar{\psi}^{I} \ .
\end{equation}
With this definition, the Grassmann version
of the eigenvector equations follows nicely as
\begin{equation}
  \int_{\bar{\psi}\psi}v^\dagger_{\psi}\mathcal{M}_{\bar{\psi}\phi}
  =\lambda v^\dagger_{\phi} \ ,
  \qquad\int_{\bar{\phi}\phi}\mathcal{M}_{\bar{\psi}\phi}u_{\bar{\phi}}
  =\lambda u_{\bar{\psi}} \ .
\end{equation}
Next the Hermitian conjugate of $\mathcal{M}$ is given by
\begin{equation}
  \mathcal{M}^\dagger_{\bar{\psi}\phi}
  =\sum_{I,J}M^{(\text{m})*}_{JI}\sigma_{I}\bar{\psi}^{I}{\phi}^{J} \ .
  \label{eq:conjugate}
\end{equation}
Note how the notion of Hermiticity, unitarity, and duality
of the vector space is straightforward and intuitive when
the coefficient tensor is written in the matrix format.
With all of these defined, we can write the Grassmann version of the
singular value decomposition (gSVD) or the eigen decomposition (gEigD) as
\begin{equation}
    \mathcal{M}_{{\bar\psi}\phi}=\int_{\bar{\eta}\eta}\int_{\bar{\zeta}\zeta}
    \mathcal{U}_{\bar{\psi}\eta}
    \text{diag}(\lambda)_{\bar{\eta}\zeta}
    \mathcal{V}^\dagger_{\bar{\zeta}\phi} \ ,
\end{equation}
where
\begin{equation}
    \text{diag}(\lambda)_{\bar{\phi}\psi}=\sum_{I}\lambda_{I}\sigma_{I}\bar{\phi}^{I}{\psi}^{I} \ .
\end{equation}
The unitary matrices $\mathcal{U}$, $\mathcal{V}$
and $\text{diag}(\lambda)$ are obtained as follows.
We first obtain the coefficient matrix $M^\text{(m)}_{IJ}$
with the parity-preserving encoder \eqref{eq:parity_preserving_encoder}.
If $\mathcal{M}$ is Grassmann even, $M^\text{(m)}$ can be
diagonalized into two blocks with $I,J$ even and odd, respectively. 
\begin{equation}
  M^\text{E}_{IJ}
  =M^{(\text{m})}_{2I,2J},\quad M^\text{O}_{IJ}=M^{(\text{m})}_{2I-1,2J-1} \ .
    \label{eq:Mblocks}
\end{equation}
These two blocks are diagonalized separately and
give two sets of unitary matrices
$(U^\text{E},V^\text{E})$ and $(U^\text{O},V^\text{O})$.
The full unitary matrices $(U^{(\text{m})},V^{(\text{m})})$ are
subsequently obtained by
\begin{align}
  U^{(\text{m})}_{2I,2J}&=U
  ^\text{E}_{IJ},\quad U^{(\text{m})}_{2I-1,2J-1}=U^\text{O}_{IJ} \ ,
    \label{eq:Ublocks}\\
    V^{(\text{m})}_{2I,2J}&=
    V^\text{E}_{IJ},\quad V^{(\text{m})}_{2I-1,2J-1}=V^\text{O}_{IJ} \ .
    \label{eq:Vblocks}
\end{align}
Notice that the two matrix indices of $U^{(m)}$ and $V^{(m)}$ also have
the same parity, which means that the Grassmann isometry
$\mathcal{U}$ obtained from $U^\text{(m)}$ is guaranteed to be Grassmann even,
and so is $\mathcal{V}$. Because all the matrices in the decomposition are
Grassmann even, all the tensors formed and decomposed during
the coarse-graining procedures are also always Grassmann even,
so the process is free from the global sign factor when we exchange the
position of the tensors.

\section{Details of the coarse-graining procedures}
\label{section:algorithm}

In this section, we present the details of
the coarse-graining procedures.
There are actually two kinds of
them used in this paper,
namely the HOTRG method, which
is used when we block the tensor in the $z$(flavor)-direction,
and
the TRG method, which is used in two-dimensional blocking.

\subsection{Grassmann HOTRG}
\label{section:HOTRG}

\begin{figure}
    \centering
    \includegraphics{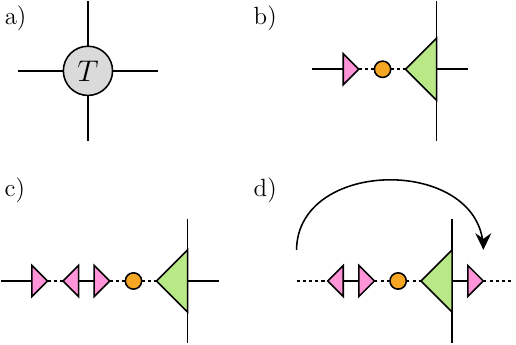}
    \caption{Schematic representation of \eqref{eq:T_SVD}-\eqref{eq:T_SVD2}.
      Small triangles, large triangles, and circles represent
      $U_{ia}$, $V_{ajkl}$ and $\lambda_a$, respectively.
      The dashed lines represent
      the truncated legs. a) A four-legged tensor $T_{ijkl}$.
      b) An SVD is performed for one of the legs, $i$.
      c) A resolution of the identity $U^\dagger U$ is inserted
      between $U$ and $\lambda$.
      d) An isometry is moved to the other side.}
    \label{fig:hotrg_concept}
\end{figure}

Let us first give a brief review of the non-Grassmann HOTRG method.
The main idea is
to insert a resolution of the identity on the bonds we wish to truncate.
For example, given a two-dimensional tensor $T_{ijkl}$, where
we wish to truncate the leg $i$,
the most naive way is to perform an SVD and insert
an identity $U^\dagger U$ on that leg as
\begin{equation}
    T_{ijkl}=\sum_aU_{ia}\lambda_aV_{ajkl}=\sum_{a,b,i'}U_{ib}(U^\dagger_{bi'}U_{i'a})\lambda_aV_{ajkl} \ ,
    \label{eq:T_SVD}
\end{equation}
where the legs $a$ and $b$ are truncated.
The schematic representation of this process is shown in
Fig.~\ref{fig:hotrg_concept}-a to -c.
If the legs $i$ and $k$ in $T_{ijkl}$ are contracted on a periodic lattice,
which means that they are
pointing in the opposite direction,
the isometry $U_{ib}$ can be moved to contract with $k$ instead as
\begin{equation}
    \sum_{a,b,i'}U_{ib}U^\dagger_{bi'}U_{i'a}\lambda_aV_{ajkl}\rightarrow
    \sum_{a,i,k}U^\dagger_{bi}U_{ia}\lambda_aV_{ajkl}U_{kc}=
    \sum_{i,k}U^\dagger_{bi}T_{ijkl}U_{kc}\equiv T'_{bjcl} \ .
    \label{eq:T_SVD2}
\end{equation}
Since the legs $b$ and $c$ are truncated, the tensor $T'$ becomes smaller.
This process is shown in Fig.~\ref{fig:hotrg_concept}-c to -d.
Note that the isometry is not unique. If we apply the same procedure on
the leg $k$ instead, we will arrive at a different isometry.
We have to choose the one that makes the singular value spectrum
decay the fastest and thus allows the smallest truncation.

A more efficient but equivalent way to do this is to define a Hermitian matrix
\begin{equation}
    M_{ii'}=\sum_{j,k,l}T_{ijkl}T^*_{i'jkl} \ .
    \label{eq:M_SVD}
\end{equation}
Substituting the first equality of \eqref{eq:T_SVD} in \eqref{eq:M_SVD},
we get
\begin{equation}
    M_{ii'}=\sum_{j,k,l,a,b}U_{ia}\lambda_aV_{ajkl}U^*_{i'b}\lambda_bV^*_{bjkl}=\sum_{j,k,l,a,b}U_{ia}\lambda_a^2U^\dagger_{ai'} \ .
\end{equation}
In other words, we can obtain $U$ by diagonalizing the matrix $M$
instead of performing an SVD for a four-legged tensor $T$, which is much slower.

We can straightforwardly apply such a coarse-graining procedure to the 3-dimensional case.
Given a 6-leg tensor $T_{i_1i_2i_3j_1j_2j_3}$ with $i_\mu$ and $j_\mu$ pointing in the opposite
direction, we first contract two $T$ tensors in $z$ direction,
\begin{equation}
  \sum_{k_3}T_{i_1i_2i_3j_1j_2k_3}T_{i'_1i'_2k_3j'_1j'_2j_3}
  = \tilde T_{(i_1i'_1)(i_2i'_2)i_3(j_1j'_1)(j_2j'_2)j_3} \ .
\end{equation}
and then attach the isometry $(U_\mu)_{(ii')\tilde i}$ to merge the double bond $(i_\mu i'_\mu)$ into a truncated
bond $\tilde i_\mu$ . Note that the standard HOTRG procedure can be generalized to the Grassmann
tensor network \cite{Akiyama:2020sfo,Bloch:2022vqz}.

\begin{figure}
    \centering
    \includegraphics[scale=0.8]{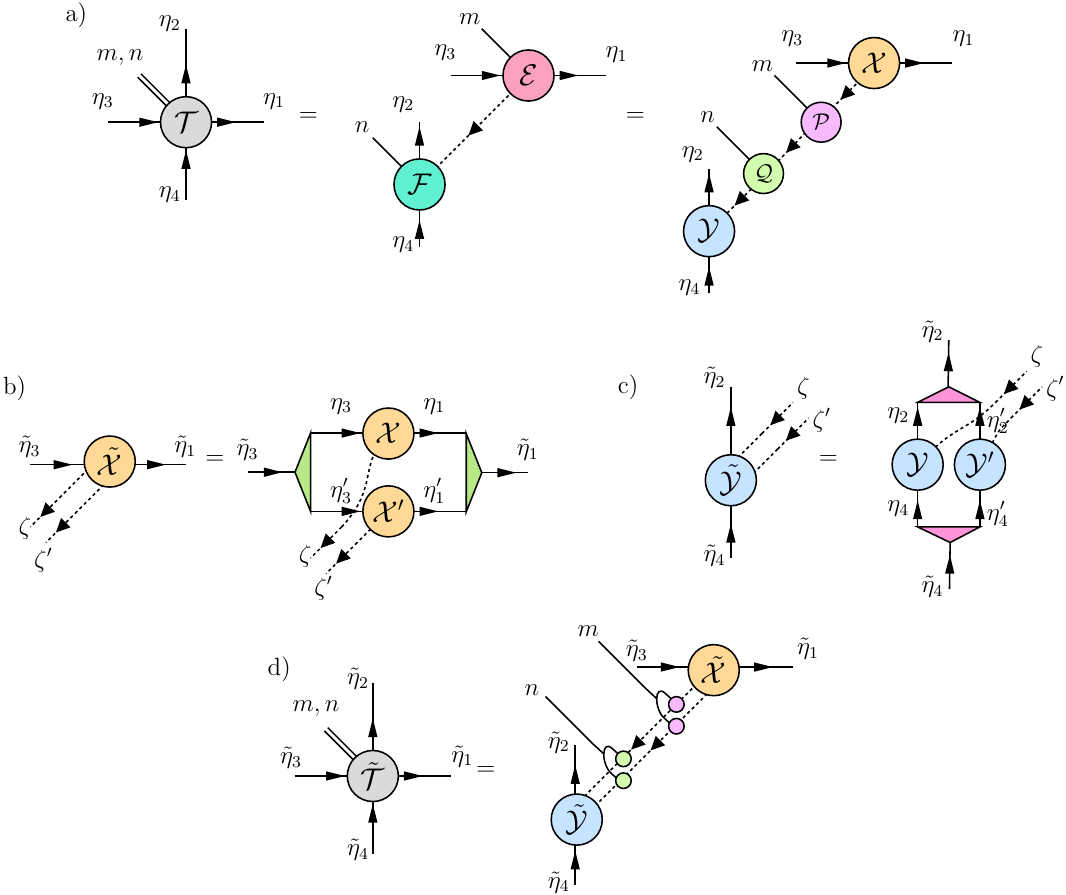}
    \caption{Schematic representation of the Grassmann HOTRG algorithm
      for the site tensor \eqref{eq:compressdT}.
      a) the $\mathcal{X}$-$\mathcal{Y}$ decomposition \eqref{eq:TintoXY},
      b) $\tilde{\mathcal X}$ \eqref{eq:Xtilde},
      c) $\tilde{\mathcal Y}$ \eqref{eq:Ytilde} and
      d) the coarse-grained tensor $\tilde{\mathcal T}$ \eqref{eq:Ttilde}.
      }
    \label{fig:hotrg_diagrams}
\end{figure}

For the flavor coarse-graining, however, we explain a slightly modified version of the
Grassmann HOTRG to improve computational efficiency in dealing with large-size initial
tensors. The key point is first to decompose the tensor of each layer into small sub-tensors
and then block the legs of subtensors from the layers $\alpha$ and $\alpha'$ in the $x$ and $y$ directions
separately. Namely, we perform gSVD for $\mathcal{T}^{(\alpha)}$
(See Fig.~\ref{fig:hotrg_diagrams}-a.)
\begin{align}
    \mathcal{T}^{(\alpha)}_{\eta_1\eta_2\bar\eta_3\bar\eta_4;mn}&=
    \int_{\bar\psi\psi}
    \mathcal{E}^{(\alpha)}_{\eta_1\bar\eta_3\psi;m}
    \mathcal{F}^{(\alpha)}_{\bar\psi\eta_2\bar\eta_4;n} \ ,
    \label{eq:EFdecomposition}\\
    &=\int_{\bar\zeta\zeta,\bar\phi\phi,\bar\xi\xi}
    \mathcal{X}^{(\alpha)}_{\eta_1\bar\eta_3\zeta}
    \mathcal{P}^{(\alpha)}_{\bar\zeta\phi;m}
    \mathcal{Q}^{(\alpha)}_{\bar\phi\xi;n}
    \mathcal{Y}^{(\alpha)}_{\bar\xi\eta_2\bar\eta_4} \ .
    \label{eq:TintoXY}
\end{align}
In \eqref{eq:EFdecomposition},
both $\mathcal{E}$ and $\mathcal{F}$
absorb the square root of the singular value in their definition.
In \eqref{eq:TintoXY}, $\mathcal{E}$ is further decomposed
into $\mathcal{X}$ and $\mathcal{P}$ with $\mathcal{X}$
absorbing the singular values and $\mathcal{F}$ is also decomposed
into $\mathcal{Y}$ and $\mathcal{Q}$ with $\mathcal{Y}$
absorbing the singular values.

To merge the legs from the layer $\alpha$ and $\alpha'$, we next define
\begin{align}
    \hat{\mathcal{X}}_{\bar\eta_3\bar\eta_3'\zeta\zeta'\eta_1\eta_1'}
    &=
    \mathcal{X}^{(\alpha)}_{\eta_1\bar\eta_3\zeta}
    \mathcal{X}^{(\alpha')}_{\eta'_1\bar\eta'_3\zeta'} \ ,\\
    \hat{\mathcal{Y}}_{\bar\eta_4\bar\eta_4'\bar\xi\bar\xi'\eta_2\eta_2'}
    &=
    \mathcal{Y}^{(\alpha)}_{\bar\xi\eta_2\bar\eta_4}
    \mathcal{Y}^{(\alpha')}_{\bar\xi'\eta'_2\bar\eta'_4} \ .
\end{align}
Then we form the Hermitian matrices for the isometry computation:
\begin{align}
  (\mathcal{M}_1)_{(\bar\eta_1\bar\eta_1')(\eta_1\eta_1')}&=
  \int_{\bar\eta_3\eta_3,\bar\eta'_3\eta'_3,\bar\zeta\zeta,\bar\zeta'\zeta'}
    \hat{\mathcal{X}}^\dagger_{(\bar\eta_1\bar\eta_1')(\eta_3\eta_3'\bar\zeta\bar\zeta')}
    \hat{\mathcal{X}}_{(\bar\eta_3\bar\eta_3'\zeta\zeta')(\eta_1\eta_1')} \ ,\\
    (\mathcal{M}_3)_{(\bar\eta_3\bar\eta_3')(\eta_3\eta_3')}&=\int_{\bar\eta_1\eta_1,\bar\eta'_1\eta'_1,\bar\zeta\zeta,\bar\zeta'\zeta'}
    \hat{\mathcal{X}}_{(\bar\eta_3\bar\eta_3')(\zeta\zeta'\eta_1\eta_1')}
    \hat{\mathcal{X}}^\dagger_{(\bar\zeta\bar\zeta'\bar\eta_1\bar\eta_1')(\eta_3\eta_3')} \ ,\\
    (\mathcal{M}_2)_{(\bar\eta_2\bar\eta_2')(\eta_2\eta_2')}&=\int_{\bar\eta_4\eta_4,\bar\eta'_4\eta'_4,\bar\xi\xi,\bar\xi'\xi'}
    \hat{\mathcal{Y}}^\dagger_{(\bar\eta_2\bar\eta_2')(\eta_4\eta_4'\xi\xi')}
    \hat{\mathcal{Y}}_{(\bar\eta_4\bar\eta_4'\bar\xi\bar\xi')(\eta_2\eta_2')} \ ,\\
    (\mathcal{M}_4)_{(\bar\eta_4\bar\eta_4')(\eta_3\eta_4')}&=\int_{\bar\eta_2\eta_2,\bar\eta'_2\eta'_2,\bar\xi\xi,\bar\xi'\xi'}
    \hat{\mathcal{Y}}_{(\bar\eta_4\bar\eta_4')(\bar\xi\bar\xi'\eta_2\eta_2')}
    \hat{\mathcal{Y}}^\dagger_{(\xi\xi'\bar\eta_2\bar\eta_2')(\eta_4\eta_4')} \ .
\end{align}
The unitary matrices along the same axis are then compared by their
singular values as explained before.
The resulting isometries $(\mathcal{U}_x)_{(\tilde\eta)(\bar\eta\bar\eta')}$
and $(\mathcal{U}_y)_{(\tilde\eta)(\bar\eta\bar\eta')}$ are then used
for the truncation
\begin{align}
    \tilde{\mathcal{X}}_{\bar{\tilde\eta}_3\zeta\zeta'\tilde{\eta}_1}&=
    \int_{\bar\eta_1\eta_1,\bar\eta'_1\eta'_1,\bar\eta_3\eta_3,\bar\eta'_3\eta'_3}
    (\mathcal{U}_x)_{(\bar{\tilde\eta}_3)(\eta_3\eta'_3)}
    \hat{\mathcal{X}}_{\bar\eta_3\bar\eta_3'\zeta\zeta'\eta_1\eta_1'}
    (\mathcal{U}_x)^\dagger_{(\bar\eta_1\bar\eta'_1)(\tilde\eta_1)} \ ,
    \label{eq:Xtilde}\\
    \tilde{\mathcal{Y}}_{\bar{\tilde\eta}_4\bar\xi\bar\xi'\tilde{\eta}_2}&=
    \int_{\bar\eta_2\eta_2,\bar\eta'_2\eta'_2,\bar\eta_4\eta_4,\bar\eta'_4\eta'_4}
    (\mathcal{U}_y)_{(\bar{\tilde\eta}_4)(\eta_4\eta'_4)}
    \hat{\mathcal{Y}}_{\bar\eta_4\bar\eta_4'\bar\xi\bar\xi'\eta_2\eta_2'}
    (\mathcal{U}_y)^\dagger_{(\bar\eta_2\bar\eta'_2)(\tilde\eta_2)} \ .
    \label{eq:Ytilde}
\end{align}
The final coarse-grained tensor can then be constructed as
\begin{equation}
    \tilde{\mathcal{T}}_{{\tilde\eta}_1{\tilde\eta}_2\bar{\tilde\eta}_3\bar{\tilde\eta}_4;mn}
    =
    \int_{\substack{\bar\zeta\zeta,\bar\phi\phi,\bar\xi\xi,\\\bar\zeta'\zeta',\bar\phi'\phi',\bar\xi'\xi'}}
    \tilde{\mathcal{X}}_{\bar{\tilde\eta}_3\zeta\zeta'\tilde{\eta}_1}
    \mathcal{P}^{(\alpha)}_{\bar\zeta\phi;m}
    \mathcal{P}^{(\alpha')}_{\bar\zeta'\phi';m}
    \mathcal{Q}^{(\alpha)}_{\bar\phi\xi;n}
    \mathcal{Q}^{(\alpha')}_{\bar\phi'\xi';n}
    \tilde{\mathcal{Y}}_{\bar{\tilde\eta}_4\bar\xi\bar\xi'\tilde{\eta}_2} \ .
    \label{eq:Ttilde}
\end{equation}
The schematic representation of $\tilde{\mathcal{X}}$, $\tilde{\mathcal{Y}}$
and $\tilde{\mathcal{T}}$ are shown in Fig.~\ref{fig:hotrg_diagrams}-b to -d. After repeating
the above gHOTRG appropriate times, we can finally obtain the flavor coarse-grained tensor
$\mathcal{T}''_{\psi_1\psi_2\bar\psi_3\bar\psi_4;mn}$ in the right-hand side of \eqref{eq:zblocking}, where we have used $\psi$ indices instead of $\tilde\eta$ for clarity.

\subsection{Grassmann TRG}
\label{section:2dTRG}

\begin{figure}
    \centering
    \includegraphics[scale=0.8]{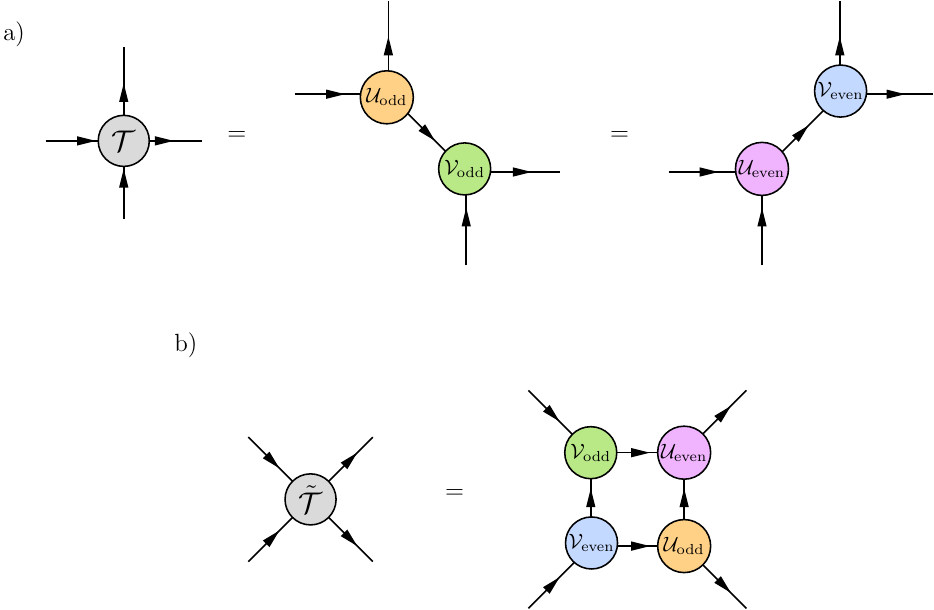}
    \caption{schematic representation of the Grassmann TRG algorithm.}
    \label{fig:trg_diagrams}
\end{figure}

As in the traditional TRG \cite{Levin:2006jai},
we perform an SVD
on $\mathcal{T}_{\eta_1\eta_2\bar\eta_3\bar\eta_4}$
in two different ways depending on whether the tensor is located on the even or odd site of a square lattice (See Fig.~\ref{fig:trg_diagrams}-a.)
as
\begin{align}
    \mathcal{T}_{\eta_1\eta_2\bar\eta_3\bar\eta_4}
    &=\int_{\bar\xi\xi}
    (\mathcal{U}_\text{odd})_{\eta_2\bar\eta_3\xi}
    (\mathcal{V}_\text{odd})_{\bar\xi\bar\eta_4\eta_1}\\
    &=\int_{\bar\xi\xi}
    (\mathcal{U}_\text{even})_{\bar\eta_3\bar\eta_4\xi}
    (\mathcal{V}_\text{even})_{\bar\xi\eta_1\eta_2} \ .
\end{align}
Practically, these gSVDs are calculated in terms of the coefficient matrix representation as
\begin{align}
    (Q_\text{odd}^\text{(m)})_{(I_2I_3)(I_4I_1)}&=T_{I_1I_2I_3I_4}\sigma_{\{I_2,I_3\}}(-)^{p(I_1)}\\
    &\overset{\text{SVD}}{=}\;
    \sum_J(U_\text{odd}^\text{(m)})_{(I_2I_3)J}(V_\text{odd}^\text{(m)})_{J(I_4I_1)} \ ,\\
    (Q_\text{even}^\text{(m)})_{(I_3I_4)(I_1I_2)}&=T_{I_1I_2I_3I_4}\sigma_{\{I_3,I_4\}}(-)^{p(I_1)+p(I_2)}\\
    &\overset{\text{SVD}}{=}\;
    \sum_J(U_\text{even}^\text{(m)})_{(I_3I_4)J}(V_\text{even}^\text{(m)})_{J(I_1I_2)} \ .
\end{align}
where the square root of singular value matrices are absorbed into $U^\text{(m)}$ and $V^\text{(m)}$. Using $U^\text{(m)}$ and $V^\text{(m)}$, one can then obtain the blocked tensor with renormalized legs as (See Fig.~\ref{fig:trg_diagrams}-b.)
\begin{equation}
    \tilde{\mathcal{T}}_{\xi_1\xi_2\bar\xi_3\bar\xi_4}=
    \int_{\bar\eta_1\eta_1,\bar\eta_2\eta_2,\bar\eta_3\eta_3,\bar\eta_4\eta_4}
    (\mathcal{U}_\text{even})_{\bar\eta_3\bar\eta_2\xi_2}
    (\mathcal{U}_\text{odd})_{\eta_2\bar\eta_1\xi_1}
    (\mathcal{V}_\text{even})_{\bar\xi_4\eta_1\eta_4}
    (\mathcal{V}_\text{odd})_{\bar\xi_3\bar\eta_4\eta_3}.
    \label{eq:TRG_Ttilde}
\end{equation}
For convenience, we also note the explicit form of the coefficient tensors below
\begin{align}
    \tilde T _{J_1J_2J_3J_4} &= \sum_{I_3I_1}U'_{I_3I_1J_1J_2}V'_{J_3J_4I_1I_3}\sigma_{I_1}\sigma_{I_3}(-)^{p(J_1)+p(J_2)} \ ,\\
    U'_{I_3I_1J_1J_2}&=\sum_{I_2}(U_\text{even}^\text{(m)})_{(I_3I_2)J_2}(U_\text{odd}^\text{(m)})_{(I_2I_1)J_1}
    \sigma_{I_2}\sigma_{\{I_3,I_2\}}\sigma_{\{I_2,I_1\}}(-)^{p(I_2)} \ , \\
    V'_{J_3J_4I_1I_3}&=\sum_{I_4}(V_\text{even}^\text{(m)})_{J_4(I_1I_4)}(V_\text{odd}^\text{(m)})_{J_3(I_4I_3)}
    \sigma_{I_4}\sigma_{J_4}\sigma_{J_3} \ .
\end{align}

\section{Adding local inter-flavor interactions}
\label{section:interflavor_interaction}

\begin{figure}
    \centering
    \includegraphics[scale=0.8]{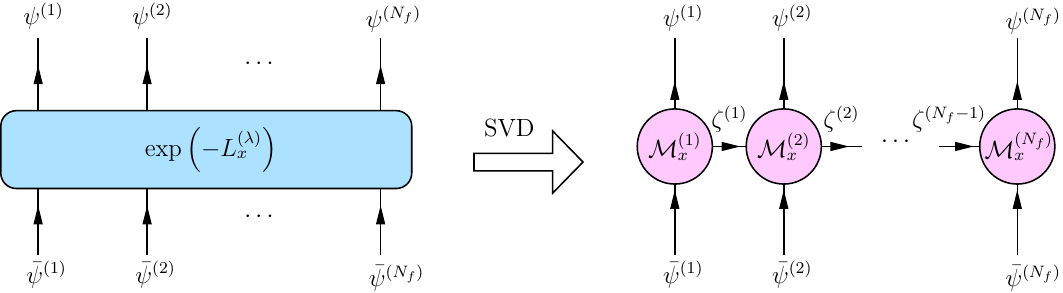}
    \caption{Decomposition of the Boltzmann weight $e^{-L_x^{(\lambda)}}$
      defined in \eqref{eq:interaction_boltzmann_weight}.}
    \label{fig:flavor_interactions_diagrams}
\end{figure}

In this paper, we have considered the case in which
different flavors interact through the gauge field only.
However, our method can be generalized to models
with local inter-flavor interactions.
Here we consider the interaction term in the Lagrangian given by
\begin{equation}
  L^{(\lambda)}_x = \sum_{I_1\cdots J_{N_{\rm f}}}\lambda_{I_1\cdots I_{N_{\rm f}}J_1\cdots J_{N_{\rm f}}}(\psi_x^{(1)})^{I_1}\cdots (\psi_x^{(N_{\rm f})})^{I_{N_{\rm f}}}(\bar\psi_x^{(1)})^{J_1}\cdots (\bar\psi_x^{(N_{\rm f})})^{J_{N_{\rm f}}}
\end{equation}
with the Boltzmann weight
\begin{equation}
    e^{-L^{(\lambda)}_x} = \sum_{I_1\cdots J_{N_{\rm f}}}B^{(\lambda)}_{I_1\cdots I_{N_{\rm f}}J_1\cdots J_{N_{\rm f}}}(\psi_x^{(1)})^{I_1}\cdots (\psi_x^{(N_{\rm f})})^{I_{N_{\rm f}}}(\bar\psi_x^{(1)})^{J_1}\cdots (\bar\psi_x^{(N_{\rm f})})^{J_{N_{\rm f}}} \ .
    \label{eq:interaction_boltzmann_weight}
\end{equation}
If we treat all the fermions as Grassmann indices,
we can perform the tensor decomposition of the Boltzmann weight
in the same spirit as in Ref.~\cite{Akiyama:2023lvr} as
\begin{equation}
    e^{-L^{(\lambda)}_x} =
    \int_{
    \substack{
    \bar\zeta^{(1)}_x\zeta^{(1)}_x,\cdots\\,\bar\zeta^{(N_{\rm f}-1)}_x\zeta^{(N_{\rm f}-1)}_x
    }}
    \mathcal{M}^{(1)}_{\psi^{(1)}_x\bar\psi^{(1)}_x\zeta^{(1)}_x}
    \mathcal{M}^{(2)}_{\bar\zeta^{(1)}_x\psi^{(2)}_x\bar\psi^{(2)}_x\zeta^{(2)}_x}
    \cdots
    \mathcal{M}^{(N_{\rm f})}_{\bar\zeta^{(N_{\rm f}-1)}_x\psi^{(N_{\rm f})}_x\bar\psi^{(N_{\rm f})}_x} \ ,
\end{equation}
as shown in Fig.~\ref{fig:flavor_interactions_diagrams}.

The partition function with this interaction term now becomes
\begin{equation}
  Z=\int_{\bar{\eta}\eta,\bar\zeta\zeta,\bar\xi\xi} \sum_{\{\varphi\}}
  \prod_{x,\alpha}\mathcal{T}_{x}^{(\alpha)}  \ ,
\end{equation}
where we have defined

\begin{align}
    \mathcal{T}_{x}^{(\alpha)}&=P^{(\alpha)}_{x} \hat{\mathcal{S}}^{(\alpha)}_{x}L^{(\alpha)}_{x,1}L^{(\alpha)}_{x,2}\mathcal{M}^{(\alpha)}_x \ ,
    \label{eq:site_tensor_original_with_interaction}\\
    \hat{\mathcal{S}}^{(\alpha)}_{x}&=
    \int d\psi_{x}^{(\alpha)}d\bar{\psi}_{x}^{(\alpha)}e^{-\bar{\psi}_{x}^{(\alpha)} W_{x}^{(\alpha)}\psi_{x}^{(\alpha)}
    -\sum_{\pm,\nu}
    \left\{\bar{\psi}_{x}^{(\alpha)}\eta_{x,\pm\nu}^{(\alpha)}-\bar{\eta}_{x\mp\hat\nu,\pm\nu}^{(\alpha)}H_{x\mp\hat\nu,\pm\nu}^{(\alpha)}\psi_{x}^{(\alpha)}\right\}}\mathcal{I}_{\bar\psi^{(\alpha)}_x\xi_x^{(\alpha)}}\mathcal{I}_{\bar\xi_x^{(\alpha)}\psi^{(\alpha)}_x} \ ,
    \label{eq:site_tensor_S_with_interaction}\\
    \mathcal{I}_{\bar\psi\xi}&=\sum_I\sigma_I\bar\psi^I\xi^I \ .
\end{align}

\begin{figure}
    \centering
    \includegraphics[scale=0.7]{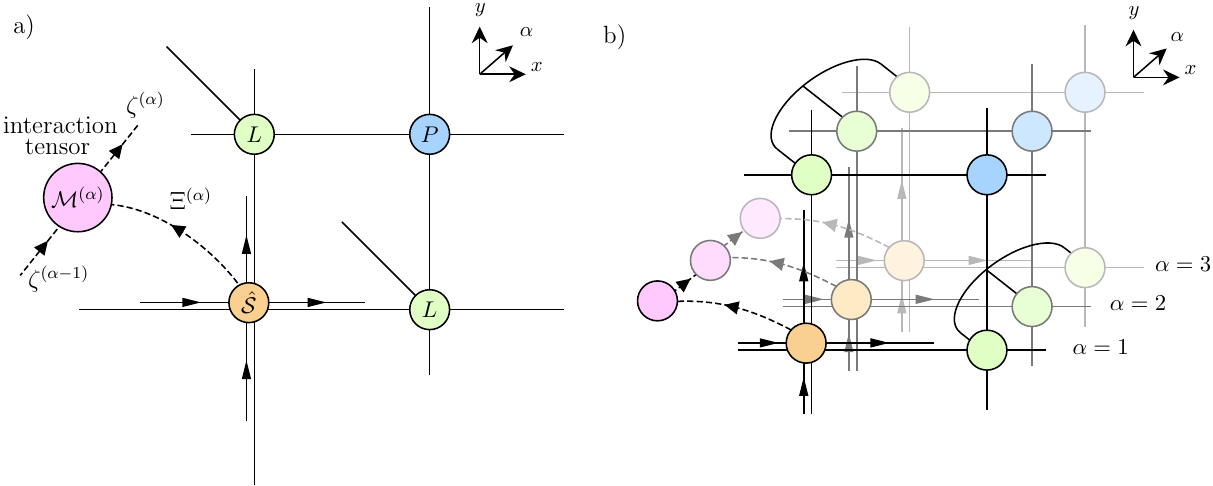}
    \caption{Connection of the tensors with the local multi-flavor interaction.
      a) The site tensor.
      b) The connection between layers corresponding to the $N_{\rm f}=3$ case.}
    \label{fig:interaction_connection}
\end{figure}

The difference between this result and the original
one in \eqref{eq:site_tensor_original}-\eqref{eq:site_tensor_S} is
that we now have the interaction tensor
$\mathcal{M}^{(\alpha)}_{\bar\zeta^{(\alpha-1)}\xi^{(\alpha)}\bar\xi^{(\alpha)}\zeta^{(\alpha)}}$,
which is connected with $\hat{\mathcal{S}}^{(\alpha)}$
through two extra connection tensors $\mathcal{I}_{\bar\psi^{(\alpha)}\xi^{(\alpha)}}$
and $\mathcal{I}_{\bar\xi^{(\alpha)}\psi^{(\alpha)}}$.
These connection tensors are actually the Grassmann identity matrices
which have the property that their contraction with any Grassmann tensor
always gives the same tensor.
The legs $(\bar\xi^{(\alpha)},\xi^{(\alpha)})$ can be merged into a single
fermion $\Xi^{(\alpha)}$ with the
prescription \eqref{eq:join_leg1}-\eqref{eq:join_leg2}.
The connection of the interaction tensor $\mathcal{M}$ with $\hat{\mathcal{S}}$ on the lattice is shown
in Fig.~\ref{fig:interaction_connection}.

Since the general structure of the tensor network is similar to that in Fig.~\ref{fig:site_tensor}, appropriate compression and coarse-graining techniques can be straightforwardly applied to this site tensor.

\bibliographystyle{JHEP}
\bibliography{ref}

\end{document}